\documentclass[3p,11pt,authoryear]{elsarticle}

\usepackage{natbib}
\usepackage{subcaption}
\usepackage{amsfonts}
\usepackage{amsmath}
\usepackage{amssymb}
\usepackage{amsthm}
\usepackage{mathrsfs}
\usepackage{mathtools}
\usepackage{graphics}
\usepackage{lscape}
\usepackage{changebar}
\usepackage{graphicx}
\usepackage{epsfig}
\usepackage{bm}
\usepackage{color}
\usepackage{hyperref}
\newcommand{\tr}{\hbox{tr}}

\newcommand{\noop}[1]{} 

\renewcommand{\=}{\!=\!}

\journal{Journal of the Mechanics and Physics of Solids}

\begin{document}

\begin{frontmatter}

\title{Oscillatory and tip-splitting instabilities in 2D dynamic fracture:\\
The roles of intrinsic material length and time scales}

\author{Aditya Vasudevan$^{1}$\footnote{Equal contribution}, Yuri Lubomirsky$^{2\,*}$, Chih-Hung Chen$^{1,3}$,\\ Eran Bouchbinder$^2$\footnote{Corresponding author. E-mail address: \url{a.karma@northeastern.edu}, \url{eran.bouchbinder@weizmann.ac.il}}, Alain Karma$^{1\,\dagger}$}
\address{$^1$Department of Physics and Center for Interdisciplinary Research on Complex Systems,\\ Northeastern University, Boston, Massachusetts 02115, USA\\
$^2$Chemical and Biological Physics Department, Weizmann Institute of Science, Rehovot 7610001, Israel\\
$^3$Institute of Applied Mechanics, National Taiwan University, Taipei 106, Taiwan}
\date{\today}
\begin{abstract}
Recent theoretical and computational progress has led to unprecedented understanding of symmetry-breaking instabilities in 2D dynamic fracture. At the heart of this progress resides the identification of two intrinsic, near crack tip length scales --- a nonlinear elastic length scale $\ell$ and a dissipation length scale $\xi$ --- that do not exist in Linear Elastic Fracture Mechanics (LEFM), the classical theory of cracks. In particular, it has been shown that at a propagation velocity $v$ of about $90\%$ of the shear wave-speed, cracks in 2D brittle materials undergo an oscillatory instability whose wavelength varies linearly with $\ell$, and at larger loading levels (corresponding to yet higher propagation velocities), a tip-splitting instability emerges, both in agreements with experiments. In this paper, using phase-field models of brittle fracture, we demonstrate the following properties of the oscillatory instability: (i) It exists also in the absence of near-tip elastic nonlinearity, i.e.~in the limit $\ell\!\to\!0$, with a wavelength determined by the dissipation length scale $\xi$. This result shows that the instability crucially depends on the existence of an intrinsic length scale associated with the breakdown of linear elasticity near crack tips, independently of whether the latter is related to nonlinear elasticity or to dissipation. (ii) It is a supercritical Hopf bifurcation, featuring a vanishing oscillations amplitude at onset. (iii) It is largely independent of the phenomenological forms of the degradation functions assumed in the phase-field framework to describe the cohesive zone, and of the velocity-dependence of the fracture energy $\Gamma(v)$ that is controlled by the dissipation time scale in the Ginzburg-Landau-type evolution equation for the phase-field. These results substantiate the universal nature of the oscillatory instability in 2D. In addition, we provide evidence indicating that the tip-splitting instability is controlled by the limiting rate of elastic energy transport inside the crack tip region. The latter is sensitive to the wave-speed inside the dissipation zone, which can be systematically varied within the phase-field approach. Finally, we describe in detail the numerical implementation scheme of the employed phase-field fracture approach, allowing its application in a broad range of materials failure problems.
\end{abstract}

\begin{keyword}
Fracture, Cracks, Instabilities, Nonlinear Mechanics, Phase-field models
\end{keyword}
\end{frontmatter}

\section{Background and motivation}
\label{sec:intro}

Materials failure, which is mainly mediated by crack propagation, is an intrinsically complex phenomenon that couples dynamic processes at length and time scales that are separated by many orders of magnitude, giving rise to a wealth of emergent behaviors. Crack initiation and dynamics are of prime fundamental and practical importance, and have been intensively studied in the last few decades~\citep{Freund.98, Broberg.99}. Despite some significant progress, our understanding of many basic aspects of fracture dynamics remains incomplete~\citep{Fineberg.99, Review.2010, Bouchbinder.14, perspective2015}. For example, it is now well established that dynamically propagating cracks universally undergo a three-dimensional (3D) micro-branching instability, where short-lived micro-cracks branch out sideways from the parent crack~\citep{Ravi-Chandar1984, Sharon1996, Fineberg.99, Livne2005}, bearing some similarities to side-branching in dendritic crystal growth during solidification~\citep{mullins1964stability,kessler1988pattern,karma2001branching,asta2009solidification}.

Significant progress has been made in relation to the solidification instability~\citep{mullins1964stability,kessler1988pattern,karma2001branching,asta2009solidification}, mainly because the dynamical evolution of the solid-liquid interface has been shown to be governed on a continuum scale by a well-defined free-boundary problem, which can be solved numerically or analytically in certain limits. In contrast --- as of yet --- we have no comparable understanding of dynamic fracture instabilities, mainly because we do not fully understand the strongly nonlinear and dissipative physics of the localized region near crack tips, where failure is taking place. In particular, we still miss a complete understanding of the roles played in failure dynamics by intrinsic material length and time scales associated with the crack tip physics, which are entirely neglected in the classical theory of cracks --- Linear Elastic Fracture Mechanics (LEFM)~\citep{Freund.98, Broberg.99}.

\begin{figure}[ht!]
\includegraphics[width=\textwidth]{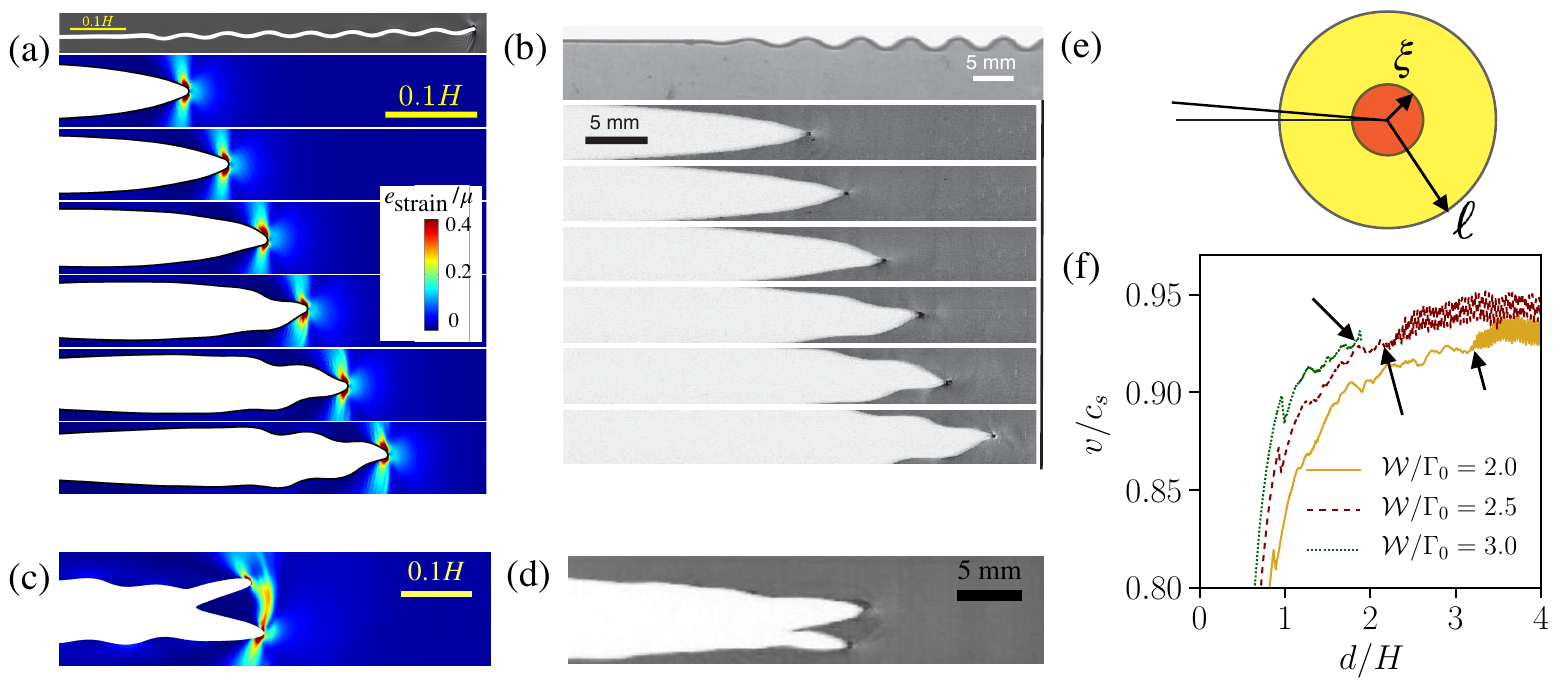}
\centering
\caption{(a) The theoretical prediction of the high-velocity oscillatory instability obtained in mode-I (tensile) phase-field fracture simulations~\citep{Chen2017,Lubomirsky2018}, see text for additional details (see also panels (b) and (f)). The top part shows the crack trajectory, defined by the $\phi\!=\!1/2$ contour (see text for details), in the undeformed configuration. The lower part presents a sequence of snapshots in the deformed configuration at the onset of instability (the color code corresponds to $e_{\mbox{\scriptsize{strain}}}/\mu$, where the elastic energy density $e_{\mbox{\scriptsize{strain}}}$ is given in Eq.~\eqref{eq:neoHookean} and $\mu$ is the shear modulus, see text for additional details. $H$ is the height of the strip in which the crack propagates). (b) The corresponding experimental images for thin brittle gels~\citep{Livne.07}, where the oscillatory instability occurs upon surpassing a critical propagation velocity of $v_{\rm c}\!\simeq\!0.92c_s$, in quantitative agreement with the theoretical-computational results of panel (a). (c) Upon increasing the driving force for fracture, crack oscillations are predicted to be followed by a tip-splitting instability (the same presentational scheme as panel (a)), see also panel (f). (d) The corresponding experimental image~\citep{Lubomirsky2018}, confirming the prediction. (e) A schematic sketch of the two intrinsic, near crack tip, material length scales discussed in this work, see extensive discussion in the text. $\ell$ is the near tip nonlinear elastic length and $\xi$ is the dissipation length. (f) The instantaneous crack velocity $v$, in units of the shear wave-speed $c_s$, as a function of the normalized propagation distance $d/H$, as obtained in large-scale phase-field simulations at different dimensionless crack driving forces ${\cal{W}}/\Gamma_0$ (see text for the definition of ${\cal W}$ and $\Gamma_0$, and the legend for the values used). For the two smallest values of ${\cal{W}}/\Gamma_0$, the crack exhibits the oscillatory instability (cf.~panels (a) and (b)) upon surpassing a critical velocity of $v_{\rm c}\!=\!0.92c_s$, marked by the arrows (note that the larger ${\cal{W}}/\Gamma_0$ is, the larger the acceleration is). For the largest ${\cal{W}}/\Gamma_0$, the crack oscillates and then tip-splits (cf.~panels (c) and (d)) at a slightly larger velocity (also marked by an arrow). The simulation parameters in panel (a) are $\Gamma_0/\mu \xi\!=\!0.29$, $H\!=\!300\xi$, $W\!=\!900\xi$, ${\cal{W}}/\Gamma_0\!=\!2.5$, $\Delta\!=\!0.21\xi$ and $\beta\!=\!0.28$ (see text and Appendix for the definition of all quantities). The parameters in panel (c) are the same, except for ${\cal{W}}/\Gamma_0\!=\!3.0$ (compare to panel (f)).}
\label{Fig:Intro-fig}
\end{figure}

Recent progress in understanding dynamic fracture instabilities has been directly related to intrinsic material length scales~\citep{Bouchbinder.08a,Livne.08,Bouchbinder.09, Bouchbinder.2009a, Livne2010,bouchbinder2010autonomy,Goldman2012, Bouchbinder.14, perspective2015,Chen2017,Lubomirsky2018}. It has been shown that a nonlinear elastic length scale $\ell$, i.e.~a length scale that is associated with nonlinear elastic deformation near the crack tip (where linear elasticity breaks down, cf.~Fig.~\ref{Fig:Intro-fig}e), controls the high velocity oscillatory instability in quasi-two-dimensional (quasi-2D) fracture, cf.~Fig.~\ref{Fig:Intro-fig}a-b for simulations and experiments, respectively. This oscillatory instability, occurring at a crack propagation velocity $v$ of $\sim\!90\%$ of the shear wave-speed $c_s$, has been experimentally observed in~\citet{Livne.07} by suppressing the 3D micro-branching instability, which typically occurs at $v\!\simeq\!0.4c_s$ or slower~\citep{Fineberg.99}, through reducing the system thickness (approaching the 2D limit). The nonlinear elastic length $\ell$ has been understood in the framework of the ``Weakly Nonlinear Elastic Theory of Fracture''~\citep{Bouchbinder.08a,Bouchbinder.09, Livne2010,bouchbinder2010autonomy,Bouchbinder.14} that extends LEFM to incorporate elastic nonlinearities near the crack tip. This theory shows that $\ell$ corresponds to a crossover between the classical square root crack tip singularity $\sim\!1/\sqrt{r}$ of LEFM, where $r$ is the distance from the tip, and a stronger $\sim\!1/r$ singularity, associated with weak elastic nonlinearities. The decisive role played by the intrinsic length scale $\ell$ in the high velocity oscillatory instability highlights basic limitations of LEFM, which features only extrinsic/geometric length scales~\citep{Bouchbinder.14}.

These new physical insights regarding the importance of near crack tip nonlinearity and intrinsic length scales have been recently incorporated into a unified theoretical and computational framework~\citep{Chen2017,Lubomirsky2018}. The latter belongs to a rather broad class of phase-field approaches to brittle fracture~\citep{karma2001phase, Karma2004, Henry2004, hakim2005crack, Henry.08, Hakim.09,aranson2000continuum,eastgate2002fracture,marconi2005diffuse,bourdin2000numerical,bourdin2008variational,bourdin2011time,miehe2010thermodynamically,ambati2015review,bleyer2017microbranching,bleyer2017dynamic,Dolbow2019,mandal2020evaluation}, which allow a self-consistent selection of the fracture-related dissipation, the crack propagation velocity $v$ and the crack path, and is particularly suitable for studying complex crack patterns under both quasi-static and dynamic conditions. Phase-field approaches have also been developed to model ductile fracture that is inherently quasi-static~\citep{ambati2015phase,miehe2016phase}. The phase-field fracture approach has proved to be highly fruitful in elucidating various material failure processes involving complex geometries --- such as crack front segmentation in mixed-mode fracture~\citep{pons2010helical,Chen2015}, quasi-static crack oscillations in thermal fracture~\citep{Corson2009}, thermal shocks~\citep{Bourdin2014} and crack kinking in anisotropic materials~\citep{Mesgarnejad2020} ---, to name a few examples. As phase-field approaches offer a self-consistent mathematical formulation of fracture problems, they inevitably also involve a dissipation length $\xi$ over which elastic singularities are regularized (cf.~Fig.~\ref{Fig:Intro-fig}e), giving rise to a well-defined $v$-dependent fracture energy $\Gamma(v)$. While the dissipation length $\xi$ in existing phase-field approaches is not associated with realistic dissipation mechanisms (e.g.~plastic deformation), but rather serves as a mathematical regularization length that mimics an effective cohesive zone and renders the fracture problem self-contained, its mere existence is completely generic. Finally, $\Gamma(v)$ also incorporates a dissipation time scale, which like the nonlinear length scale $\ell$ and the dissipation length $\xi$, is also entirely missing in LEFM.

The phase-field fracture approach of~\citet{Chen2017,Lubomirsky2018}, to be detailed below in Sec.~\ref{sec:PF}, distinguishes itself from previous approaches by incorporating near-tip elastic nonlinearities and by allowing unprecedentedly high crack propagation velocities, approaching the theoretical limiting speed (cf.~Fig.~\ref{Fig:Intro-fig}f). These novel properties resulted in a theory that predicted the high velocity 2D oscillatory instability~\citep{Chen2017}, which has been shown to be controlled by the intrinsic length scale $\ell$ (cf.~Fig.~\ref{Fig:Intro-fig}a), in quantitative agreement with experiments~\citep{Chen2017}. Furthermore, the very same theoretical and computational framework demonstrated that upon increasing the driving force for fracture ${\cal W}$, i.e.~the stored elastic energy per unit area ahead of the crack, cracks accelerate faster and to yet higher velocities (cf.~Fig.~\ref{Fig:Intro-fig}f), leading to tip-splitting after the onset of oscillations (cf.~Fig.~\ref{Fig:Intro-fig}c) or even prior to it~\citep{Lubomirsky2018}. This ultra-high velocity 2D tip-splitting instability, to be distinguished from the 3D micro-branching instability, has been then observed experimentally in the same ultra-high velocity regime~\citep{Lubomirsky2018}.

This progress gave rise to several outstanding questions that we aim at addressing in this paper. First, it has been established that the wavelength $\lambda$ of the oscillatory instability scales linearly with the nonlinear elastic length scale $\ell$ (reproduced here in Fig.~\ref{fig:LEPF-panel}a), in quantitative agreement with experiments on brittle gels~\citep{Bouchbinder.09,Goldman2012,Bouchbinder.14,Chen2017}. Extrapolating this linear relation to the $\ell\!\to\!0$ limit, i.e.~to situations in which crack tip elastic nonlinearity is small/absent, indicated a finite intercept of $\lambda\!\simeq\!13\xi$ (cf.~Fig.~\ref{fig:LEPF-panel}a). If valid, this extrapolation implies that crack tip elastic nonlinearity is not a necessary condition for the existence of the 2D oscillatory instability, i.e.~that the latter can also be controlled by the dissipation length $\xi$. In this case, the 2D oscillatory instability is entirely universal and controlled by either $\ell$ or $\xi$, i.e.~it requires an intrinsic length scale over which linear elasticity breaks down (be it nonlinear elastic or dissipative in nature) and hence is expected to be observed by any material in 2D. Yet, calculations with $\ell\=0$ did not reveal the 2D oscillatory instability, thereby suggesting instead that this instability disappears in the $\ell\!\to\!0$ limit and consequently that crack tip elastic nonlinearity is essential for its existence.

In Sec.~\ref{sec:oscillatory} we resolve this puzzle by properly probing the $\ell\!\to\!0$ limit, using a modified phase-field formulation. The latter introduces degradation functions, i.e.~functions which control the softening of the elastic energy at large strains near the crack tip, that minimize lattice (numerical grid) pinning effects. Lattice pinning effects are inherently present in any finite-difference implementation of the phase-field equations on a regular lattice/grid and tend to trap crack trajectories along lattice/grid planes, thereby suppressing small-amplitude oscillatory instabilities. We find that lattice pinning effects can be minimized by choosing degradation functions that increase the length of the effective cohesive zone along the crack propagation direction, thereby allowing us to quantitatively investigate the oscillatory instability in the $\ell\!\to\!0$ limit.  We show that the 2D oscillatory instability persists also in the absence of near-tip elastic nonlinearity, i.e.~in the limit $\ell\!\to\!0$, with a wavelength determined by the dissipation length scale $\xi$, in quantitative agreement with the linear extrapolation. This result shows that the instability crucially depends on the existence of an intrinsic length scale associated with the breakdown of linear elasticity near crack tips, independently of whether it is related to nonlinear elasticity or to dissipation.

Another open question concerns the nature of the oscillatory instability. We show that the oscillatory instability is a supercritical Hopf bifurcation, featuring a vanishing oscillations amplitude at onset. Furthermore, it remained unclear whether the oscillatory instability depends on basic properties of the phase-field model (to be introduced in detail in~Sec.~\ref{sec:PF}) including: (i) the functional form of the degradation functions that phenomenologically describe the energetic properties of the effective cohesive zone, and (ii) the dissipation time scale associated with the Ginzburg-Landau-type dynamics assumed to govern the evolution of the phase-field, which yields a velocity-dependent fracture energy $\Gamma(v)$~\citep{Karma2004}. To address the role of (i), we study rapid fracture in two phase-field formulations where the mass density in the dissipation zone is degraded as in~\citet{Chen2017,Lubomirsky2018} to attain ultra-rapid speeds. In the first, the degradation functions in the elastic energy are chosen to be the same as in the original model of Karma, Kessler and Levine (KKL)~\citep{karma2001phase}. In the second, those functions are chosen to have different forms that yield an elongated cohesive zone, which reduces the aforementioned lattice pinning effect. Therefore, this second formulation has the 2-fold benefit of allowing one to both study the onset of the oscillatory instability in the $\ell\!\to\!0$ limit and to test to what degree dynamic fracture instabilities depend on details of the energetic properties of the cohesive zone. We find that both formulations exhibit strikingly similar ``phase diagrams'' distinguishing regimes of straight, oscillatory, and tip-splitting crack states as a function of applied load and $\ell/\xi$ (the ratio of nonlinear and dissipation length scales).

To address the role of (ii), we investigate crack behavior in the KKL model as a function of the dimensionless ratio $\beta\=\tau\,c_s/\xi$ of a dissipation time scale $\tau$ and the characteristic time scale $\xi/c_s$ of energy transport on the scale $\xi$. This ratio controls the $v$ dependence of the function $\Gamma(v)$, which is nearly $v$ independent for small $\beta$, as in ideally brittle materials such as silica glass, but that becomes a steep function of $v$ for $\beta$ larger than unity, as exemplified by polymeric materials such as PMMA. We find that the onset of the oscillatory instability and its characteristic wavelength are largely independent of $\beta$ over an order of magnitude variation that encompasses the limits where $\Gamma(v)$ is weakly and strongly dependent on $v$. Taken together, the results to be presented in Sec.~\ref{sec:oscillatory} substantiate the universal nature of the oscillatory instability, which is expected to be observed in any material in 2D.

Another set of open questions concerns the physical origin of the ultra-high velocity 2D tip-splitting instability. In Sec.~\ref{sec:tip-splitting} we address this issue, where we propose that the tip-splitting instability is controlled by a limiting rate of elastic energy transport inside the crack tip region. This rate of elastic energy transport is sensitive to the wave-speed inside the dissipation zone; while the latter is not expected to change significantly compared to the elastic bulk wave-speed, it can still be systematically reduced within the phase-field approach. By so doing, we show that the critical tip-splitting velocity continuously shifts to smaller values, lending support to the proposed instability mechanism. Finally, some discussion and concluding remarks are offered in Sec.~\ref{sec:summary} and a detailed description of the numerical implementation scheme of the employed phase-field fracture approach (that is presented in Sec.~\ref{sec:PF}) appears in~\ref{sec:AppendixA} and in~\ref{sec:AppendixB}. The power of the dynamic phase-field approach is also demonstrated in~\ref{sec:AppendixC} in elucidating strongly inertial effects on the near tip fields of rapid cracks.

\section{A nonlinear phase-field approach to dynamic fracture: resolving physically-relevant, intrinsic material length scales}
\label{sec:PF}

The nonlinear phase-field approach to dynamic fracture, to be employed in this paper, has been introduced in quite some detail in~\citet{Chen2017} and studied in~\citet{Chen2017,Lubomirsky2018}. Its presentation is repeated here for completeness, and in order to further highlight its physical content and potential utility. This phase-field approach is a Lagrangian field theory that is designed to incorporate the intrinsic material length scales $\ell$ and $\xi$, and to allow for high crack propagation velocities, where dynamic instabilities are known to occur experimentally. The starting point is the Lagrangian $L\=T-U$, where the potential energy $U$ and kinetic energy $T$ are given as
\begin{eqnarray}
U\!=\!\int \left[\frac{1}{2}\kappa\left(\nabla\phi\right)^{2}+ g(\phi)\mu\,\bar{e}_{\mbox{\scriptsize{strain}}}({\bm u}) + w(\phi) e_{\rm c}\right]dV \quad\qquad\hbox{and}\quad\qquad T\!=\!\int\!\frac{1}{2}f(\phi)\rho\left(\frac{\partial {\bm u}}{\partial t}\right)^2 dV \ ,
\label{Eq:Lagrangian}
\end{eqnarray}
in terms of the displacement vector field ${\bm u}(x,y,t)$ and the scalar phase-field $\phi(x,y,t)$, an auxiliary field to be discussed below (here $(x,y)$ is a Cartesian coordinate system and $t$ is time). The elastic strain energy density functional is $e_{\mbox{\scriptsize{strain}}}({\bm u})\=\mu\,\bar{e}_{\mbox{\scriptsize{strain}}}({\bm u})$ (i.e.~$\bar{e}_{\mbox{\scriptsize{strain}}}({\bm u})$ is the dimensionless energy density functional, measured in units of the shear modulus $\mu$) and $\rho$ is the mass density. The integral are performed over the entire system and $dV$ is a volume element.

The phase-field $\phi(x,y,t)$ is a scalar field that varies continuously near the crack tip and is meant to {\em mathematically} represent the degradation process of the material upon failure. The latter process is mediated by the degradation functions $g(\phi)$, $f(\phi)$ and $w(\phi)$ that spontaneously generate, once coupled to the dissipative evolution of $\phi$, the traction-free boundary conditions on the crack faces and at the same time give rise to a finite fracture energy $\Gamma$. It is important to note that this process is a phenomenological approach
that regularizes crack tip singularities and renders the fracture problem fully self-contained, but it does not represent physically realistic dissipation processes near the crack tip. Yet, as will be further discussed below, this regularization method should satisfy some important physical constraints. Within the phase-field approach, an intact/unbroken material corresponds to $\phi\=1$, for which $g(1)\=f(1)\=1$ and $w(1)\=0$, which in turn leads to $U\=\int\!\mu\,\bar{e}_{\mbox{\scriptsize{strain}}}({\bm u})dV$ and $T\=\int\!\tfrac{1}{2}\,\rho(\partial{\bm u}/\partial t)^2 dV$. The latter correspond to an elastic material that is characterized by a linear shear wave-speed $c_s\!\equiv\!\sqrt{\mu/\rho}$, even though the elastic energy density functional $\bar{e}_{\mbox{\scriptsize{strain}}}({\bm u})$ is not necessarily quadratic, i.e.~not restricted to linear elasticity.

As an elastic material, in itself, does not contain any intrinsic length scales, we next explain how the intrinsic material length scales $\ell$ and $\xi$ are incorporated into the phase-field approach. These are related to the properties of the elastic functional $\bar{e}_{\mbox{\scriptsize{strain}}}({\bm u})$ and to the phase-field $\phi$. Consider physical situations in which the material is loaded far from the crack by weak forces, which is the generic case in brittle materials, and set $\phi\=1$. As the driving forces are weak, the material response would be predominantly linear elastic, i.e.~the quadratic approximation to $\bar{e}_{\mbox{\scriptsize{strain}}}({\bm u})$ is expected to be very good. Yet, as the crack tip is approached, the square root singularity of LEFM will build up and displacement gradients will not be necessarily small. Hence, if $\bar{e}_{\mbox{\scriptsize{strain}}}({\bm u})$ incorporates elastic nonlinearity, i.e.~contributions in the displacement gradient $\nabla{\bm u}$ that are higher order than quadratic, there will be a length scale near the crack tip where nonlinear elastic deformation becomes important. This occurs exactly at the nonlinear elastic length scale $\ell$ discussed above, which has been shown to scale as $\ell\!\sim\!\Gamma/\mu$~\citep{Bouchbinder.14}. Therefore, by using a nonlinear elastic $\bar{e}_{\mbox{\scriptsize{strain}}}({\bm u})$ and by keeping the far-field loading weak, the length scale $\ell$ is naturally incorporated into the phase-field approach, cf.~Fig.~\ref{Fig:Intro-fig}e. We note in passing that by considering strong far-field loading conditions, the very same framework makes it possible to study the fracture of soft materials~\citep{long2020fracture}, where elastic nonlinearity may be relevant at all scales. This interesting topic is not discussed in this paper.

When the crack tip is further approached, energy dissipation sets in, eventually leading to material failure, i.e.~to the loss of load-bearing capacity. This is accounted for in the phase-field approach by the field $\phi(x,y,t)$, which smoothly varies from $\phi\=1$ (intact/unbroken material) to $\phi\=0$ (fully broken material), and by the degradation functions $g(\phi)$, $f(\phi)$ and $w(\phi)$ that depend on it. The onset of dissipation is related to the strain energy density threshold $e_{\rm c}$ in Eq.~\eqref{Eq:Lagrangian}. As $\phi$ decreases from unity, $g(\phi)$ is chosen such that it decreases towards zero and $w(\phi)$ is chosen such that it increases towards unity. This process mimics the conversion of elastic strain energy into fracture energy, where the broken $\phi\=0$ phase/state becomes energetically favorable from the perspective of minimizing $U$ in Eq.~\eqref{Eq:Lagrangian}. For $\phi\=0$, we set $g(0)\=0$, implying that the effective shear modulus $g(\phi)\mu$ in Eq.~\eqref{Eq:Lagrangian} vanishes, i.e.~that the material lost its load-bearing capacity and traction-free boundary conditions are achieved. This process is associated with a length scale, which emerges from the combination of the energetic penalty of developing $\phi$ gradients, as accounted for by the first contribution to $U$ in Eq.~\eqref{Eq:Lagrangian} that is proportional to $\kappa$, and the $\phi$-dependent elastic energy density threshold for failure $(1-w(\phi))e_{\rm c}$ (the $\phi\=0$ state becomes energetically favored when the degraded elastic energy density $g(\phi)\mu\,\bar{e}_{\mbox{\scriptsize{strain}}}({\bm u})$ exceeds this threshold). Consequently, the characteristic length scale is $\xi\!\equiv\!\sqrt{\kappa/2e_{\rm c}}$, setting the size of the dissipation zone near the tip, cf.~Fig.~\ref{Fig:Intro-fig}e. To see the explicit connection between $\xi$ and fracture-related dissipation, note that the fracture energy at onset, $\Gamma_0\!\equiv\!\Gamma(v\!\to\!0)$, can be expressed as $\Gamma_0\= 4 e_{\rm c} \xi \int_0^1\!\sqrt{w(\phi)}\,d\phi$~\citep{karma2001phase,Hakim.09}.

In the KKL model~\citep{karma2001phase}, the same function $g(\phi)$ was used to represent both the degradation of the elastic modulus and the $\phi$-dependent threshold for failure, corresponding to $w(\phi)\=1-g(\phi)$ in the notation of Eq.~\eqref{Eq:Lagrangian}, while in the phase-field models introduced in the mathematical literature (see~\citet{bourdin2000numerical} and references therein), the function $w(\phi)$ has typically been chosen independently of $g(\phi)$, where $e_{\rm c}\,w(\phi)$ represents a mathematical regularization of the fracture energy. While various choices of degradation functions can yield a finite fracture energy in the $\xi\!\rightarrow\!0$ limit, the extra freedom to choose $w(\phi)$ independently of $g(\phi)$ offers additional benefits, such as the ability to vary the effective size of the cohesive zone~\citep{Dolbow2019}.

This freedom is exploited here with the particular choice $g(\phi)\=\phi^4$ and $ w(\phi)\=1-\phi $, which is found to increases the size of the effective cohesive zone in comparison to KKL in a way that substantially reduces numerical lattice/grid pinning effects. Pinning originates from the fact that the material displacement field becomes discontinuous on the lattice/grid scale in the fully broken region behind the crack tip. In a finite-difference discretization of the phase-field model on a periodic lattice, pinning tends to trap cracks along lattice planes. While this effect is minimized in finite element implementations that use unstructured grids, it is not completely eliminated. In the standard finite-difference implementation of phase-field models on 2D square lattices used here and in several previous studies~\citep{karma2001phase, Karma2004, Henry2004,hakim2005crack,Henry.08,Chen2017,Lubomirsky2018}, we find that pinning can be reduced by the aforementioned choice of degradation functions that, by effectively elongating the cohesive zone, pushes the displacement discontinuity further behind the crack tip on the scale $\xi$. (The tip can be defined arbitrarily as the most advanced point on the $\phi\=1/2$ contour). This turns out to be important to demonstrate the existence of the oscillatory instability in the $\ell\!\to\!0$ limit, which is presumably suppressed by lattice pinning in the KKL model, where the displacement discontinuity forms closer to the crack tip. For finite $\ell/\xi$ values, elastic nonlinearity promotes the oscillatory instability in such a way that lattice pinning is insufficient to suppress the instability in the KKL model~\citep{Chen2017,Lubomirsky2018}. In this setting, the comparison of crack behavior in the KKL model and the modified model with reduced pinning serves the different purpose of probing universal aspects of dynamical instabilities that are independent of details of the choice of degradation functions.

The intrinsic material length scales $\ell$ and $\xi$, as explained above, are incorporated into the potential energy $U$ in Eq.~\eqref{Eq:Lagrangian} through elastic nonlinearity in $\bar{e}_{\mbox{\scriptsize{strain}}}$ and through the phase-field $\phi$, respectively. The phase-field $\phi$ also appears in the kinetic energy $T$ in Eq.~\eqref{Eq:Lagrangian}, through the degradation function $f(\phi)$. What physical considerations should be taken into account in selecting $f(\phi)$? How is it related to $g(\phi)$? As explained above, $g(\phi)$ accounts for the degradation of the effective shear modulus $g(\phi)\mu$ inside the dissipation zone, which enforces the physical traction-free boundary conditions on the crack faces. Yet, this elastic modulus degradation process does not realistically represent dissipative processes near crack tips, e.g.~plastic deformation, that do not involve significant softening of elastic moduli. This, in turn, implies that the wave-speeds inside the dissipation zone are not significantly different from their elastic bulk values. Therefore, we write the kinetic energy inside the dissipation zone, i.e.~for $0\!\le\!\phi\!<\!1$, as $T\=\int\!\tfrac{1}{2}\,g(\phi)\mu\,[c_{\rm pz}(\phi)]^{-2}(\partial{\bm u}/\partial t)^2 dV$, where we defined the modified shear wave-speed $c_{\rm pz}(\phi)$ as
\begin{equation}
\label{Eq:waveSpeedxi}
c_{\rm{pz}}(\phi) \equiv \sqrt{\frac{g(\phi)\mu}{f(\phi)\rho}} = c_s\sqrt{\frac{g(\phi)}{f(\phi)}} \qquad\qquad\hbox{for}\qquad\qquad 0\le\phi<1  \ ,
\end{equation}
with $f(\phi)\rho$ being the effective mass density, and demand $c_{\rm{pz}}(\phi)\!\approx\!c_s$ (`pz' stands for `process zone', a common term for the dissipation zone in the fracture mechanics literature~\citep{Lawn.93}). The latter implies $f(\phi)\!\approx\!g(\phi)$. In~\citet{Chen2017,Lubomirsky2018}, as well as in Sec.~\ref{sec:oscillatory} below, $f(\phi)\=g(\phi)$ is used, implying that the mass density degrades inside the dissipation zone similarly to the shear modulus. This is quite different from earlier works~\citep{karma2001phase,Hakim.09} that used $f(\phi)\=1$. The implications of such choices on crack dynamics will be discussed in Sec.~\ref{sec:tip-splitting}.

Since fracture is a non-conservative phenomenon, the Lagrangian of Eq.~\eqref{Eq:Lagrangian} must be supplemented with a dissipation function, which is directly related to the phase-field $\phi$. We define the dissipation function $D$ as
\begin{equation}
D \equiv \frac{1}{2\chi}\int \left(\frac{\partial\phi}{\partial t}\right)^{2}dV \ ,
\label{Eq:dissipation}
\end{equation}
where $\chi$ is a dissipation rate coefficient, to be related below to the $v$-dependence of the fracture energy $\Gamma(v)$. The evolution of $\phi$ and ${\bm u}$ is derived from Lagrange's equations
\begin{eqnarray}
\frac{\partial}{\partial t}\left[\frac{\delta L}{\delta\left(\partial\psi/\partial t\right)}\right]-\frac{\delta L}{\delta\psi}
+\frac{\delta D}{\delta\left(\partial\psi/\partial t\right)}=0 \ ,
\label{Eq:Lagrange}
\end{eqnarray}
where $\psi=(\phi,u_x,u_y)$ (here ${\bm u}\=(u_x,u_y)$). Using Eqs.~\eqref{Eq:Lagrangian} and~\eqref{Eq:dissipation},
one obtains
\begin{eqnarray}
\frac{1}{\chi}\frac{\partial\phi}{\partial t} &=& -\frac{\delta\,{\cal U}({\bm u},\phi)}{\delta\phi}+\frac{1}{2}\,\rho\,\frac{\partial f}{\partial\phi}\,\frac{\partial {\bm u}}{\partial t}\!\cdot\!\frac{\partial {\bm u}}{\partial t} \ ,\label{eq:eom-phi}\\
\rho\,f\,\frac{\partial^{2}u_x}{\partial t^{2}}&=&-\frac{\delta\,{\cal U}({\bm u},\phi)}{\delta u_x}-\rho\,\frac{\partial f}{\partial t}\,\frac{\partial u_x}{\partial t} \ ,\label{eq:eom-ux}\\
\rho\,f\,\frac{\partial^{2}u_y}{\partial t^{2}}&=&-\frac{\delta\,{\cal U}({\bm u},\phi)}{\delta u_y}-\rho\,\frac{\partial f}{\partial t}\,\frac{\partial u_y}{\partial t} \ ,\label{eq:eom-uy}
\end{eqnarray}
where ${\cal U}({\bm u},\phi)$ is the potential energy density, $U\=\!\int{\cal U}({\bm u},\phi)dV$, which can be identified using Eq.~\eqref{Eq:Lagrangian}.

Equations~\eqref{eq:eom-phi}-\eqref{eq:eom-uy}, together with Eqs.~\eqref{Eq:Lagrangian} and~\eqref{Eq:dissipation}, can be used to calculate the time rate of change of the total energy of the system, leading to $d(T + U)/dt\=-2D\!\le\!0$~\citep{Chen2017}. The latter shows that the system follows gradient flow dynamics, where $D$ indeed accounts for the rate of dissipation, which is localized near the crack tip (where $\phi$ varies). This dissipation localization can be immediately employed to calculate the fracture energy $\Gamma(v)$. In particular, it suggests $v(\Gamma(v)-\Gamma_0)\=2D$, which upon using Eq.~\eqref{Eq:dissipation} and assuming steady-state crack propagation along the positive $x$-direction (implying $\partial_t\=-v\partial_x$), leads to \citep{Hakim.09}
\begin{equation}
\Gamma(v)=\Gamma_0+\chi^{-1}\,v\!\int \left(\frac{\partial\phi}{\partial{x}}\right)^2 dV \ .
\label{eq:v-Gamma}
\end{equation}
This result shows that the dissipation coefficient $\chi$ is responsible for the $v$-dependence of $\Gamma(v)$, which is also affected by the spatial distribution and extent of $\phi$ gradients near the tip. Finally, note that crack healing in the phase-field approach is prevented by using the irreversibility condition $\partial\phi/\partial{t}\!\le\!0$.

To conclude the presentation of the nonlinear phase-field approach to dynamic fracture, we discuss a natural way to nondimensionalize Eqs.~\eqref{eq:eom-phi}-\eqref{eq:eom-uy} and list the dimensionless groups of parameters that control them. Boundary conditions, and specific choices of the degradation functions $g(\phi)$, $f(\phi)$ and $w(\phi)$ (whose generic properties have already been discussed above), will be discussed later. A natural spatial scale is obviously $\xi$ (recall that it is given by $\xi\=\sqrt{\kappa/2e_{\rm c}}$), which sets the length unit. A natural time scale would be associated with the dissipation rate $\chi$, taking the form $\tau\!\equiv\!(2\chi e_{\rm c})^{-1}$, which sets the time unit. Energy density would be naturally measured in units of the shear modulus $\mu$ and the mass density $\rho$ would be naturally measured in units of $\mu/c_s^2$, where $c_s$ is the shear wave-speed introduced above. With these at hand, Eqs.~\eqref{eq:eom-phi}-\eqref{eq:eom-uy} can be fully nondimensionalized. The dimensionless set of equations depends on a small number of dimensionless groups of physical parameters. First, the dimensionless energy density functional $\bar{e}_{\mbox{\scriptsize{strain}}}({\bm u})$, which depends on the dimensionless displacement gradient tensor $\nabla {\bm u}$, contains elastic constants --- in the most general case both linear and nonlinear ones --- that are expressed in units of $\mu$. Second, the dimensionless set of equations depends on $e_{\rm c}/\mu$, which quantifies the ratio between the dissipation onset threshold $e_{\rm c}$ and a characteristic elastic modulus. Third, the equations depend on $\beta\=\tau\,c_s/\xi$ (already defined above), which quantifies the relative importance of material inertia and dissipation. As $\beta\!\sim\!\chi^{-1}$, it directly controls the $v$-dependence of the fracture energy according to Eq.~\eqref{eq:v-Gamma}, as will be further discussed in Sec.~\ref{sec:oscillatory}.

The ratio of the two fundamental length scales discussed in this paper, $\ell/\xi$, depends on all of these dimensionless parameters; the dimensionless nonlinear elastic constants appear in the prefactor of $\ell/\xi\!\sim\!\Gamma(v)/\mu\xi$~\citep{Bouchbinder.08a,Bouchbinder.14}. This prefactor vanishes in the absence of elastic nonlinearity, as will be discussed in Sec.~\ref{sec:oscillatory}, and in general depends also on $v/c_s$~\citep{Bouchbinder.08a,Bouchbinder.14}. $\ell/\xi$ is proportional to $e_{\rm c}/\mu$, but also depends on $\beta$ through the $v$-dependence of $\Gamma(v)$ (cf.~Eq.~\eqref{eq:v-Gamma}). These dependencies provide unprecedented control of a ratio of two intrinsic material length scales that are entirely missing in LEFM. This unique power of the phase-field approach to dynamic fracture has already proven essential in the discoveries reported in~\citet{Chen2017} and~\citet{Lubomirsky2018}, and will be further utilized in this paper. Finally, solutions of the dimensionless version of Eqs.~\eqref{eq:eom-phi}-\eqref{eq:eom-uy}, which are strongly nonlinear coupled partial differential equations, generally require large-scale numerical simulations. We provide a comprehensive description of the relevant numerical procedures and techniques in the Appendices.

\section{The oscillatory instability: The $\ell\!\to\!0$ limit, supercritical Hopf bifurcation and independence of $\Gamma(v)$}
\label{sec:oscillatory}

One of the major achievements of the phase-field approach presented in the previous section is related to the high-velocity 2D oscillatory instability, shown in~Fig.~\ref{Fig:Intro-fig}a-b and briefly discussed earlier in Sec.~\ref{sec:intro}. To apply the phase-field framework to a given physical problem, one needs to specify the relevant elastic strain energy density functional $e_{\mbox{\scriptsize{strain}}}$, the degradation functions $g(\phi)$, $f(\phi)$ and $w(\phi)$, the system's geometry and the applied boundary conditions. As the 2D oscillatory instability has been observed in thin brittle gels~\citep{Livne.07,Goldman2012,Bouchbinder.14}, whose near crack tip deformation is known to be described by 2D incompressible neo-Hookean elasticity~\citep{Livne2005,Livne.07}, we focus here on the latter that takes the form~\citep{Knowles.1983}
\begin{eqnarray}
e_{\mbox{\scriptsize{strain}}}=\mu\,\bar{e}_{\mbox{\scriptsize{strain}}}=\frac{\mu}{2}\,\left(F_{ij}F_{ij}+[\det({\bm F})]^{-2}-3\right) \ .
\label{eq:neoHookean}
\end{eqnarray}
Here ${\bm{F}}$ is the deformation gradient tensor, whose components are related to the displacement field ${\bm u}$ according to $F_{ij}\!=\!\delta_{ij}+\partial_{j}u_{i}$, where $i,j\!=\!\{x,y\}$. The nonlinearity in this energy density functional is contained inside the out-of-plane stretch ratio $[\det({\bm F})]^{-1}$, where nonlinear elastic coefficients (in units of $\mu$) can be obtained by a systematic expansion in the displacement gradient tensor $\nabla{\bm u}$~\citep{Bouchbinder.08a,Bouchbinder.14}. As explained above, this near tip nonlinearity, combined with weak far-field loading, gives rise to the existence of a finite nonlinear elastic length $\ell$.

For all the computations, except those that require reducing lattice pinning to study the $\ell=0$ limit, the degradation functions $g(\phi)$ and $w(\phi)$ are chosen following the well-studied KKL model~\citep{karma2001phase,Hakim.09} to be $g(\phi)\=4\phi^3-3\phi^4$ and $w(\phi)\=1-g(\phi)$ (see some additional discussion of this choice in~\citet{karma2001phase,Hakim.09} and note that the general properties $g(1)\=w(0)\=1$ and $g(0)\=w(1)\=0$, discussed in Sec.~\ref{sec:PF}, are satisfied). These degradation functions also satisfy $g'(0)\=w'(0)\= g'(1)\=w'(1)\=0$, which automatically limits the value of the phase-field to reside between $0$ and $1$. For $f(\phi)$, we choose $f(\phi)\=g(\phi)$~\citep{Chen2017,Lubomirsky2018}, which according to the discussion in Sec.~\ref{sec:PF}, leads to $c_{\rm pz}\=c_s$ in Eq.~\eqref{Eq:waveSpeedxi}. This choice, in contrast to the previously employed relation $f(\phi)\=1$~\citep{karma2001phase,Hakim.09}, allows probing the high-velocity regime in which the oscillatory instability has been observed experimentally~\citep{Livne.07}, as mentioned in  Sec.~\ref{sec:PF} and as will be discussed in detail in Sec.~\ref{sec:tip-splitting}.

Finally, we consider mode-I (tensile) cracks initially propagating along the symmetry line ($y\=0$, the propagation is in the positive $x$-direction) of a long strip of height $H$ (the long strip condition is mimicked by a finite strip of length $W$ using a treadmill procedure, as explained in~\ref{subsec:treadmill}). Fixed tensile displacements $u_y(\pm H/2)\=\pm\delta_y$ are imposed on the top and bottom boundaries of the strip, with $\delta_y\!\ll\!H$. The latter ensures weak loading conditions, i.e.~that the material behaves linearly elastically everywhere except for a small region of typical size $\ell$ near the tip. The driving force for fracture is quantified by ${\cal W}$, which equals to the elastic energy density associated with a uniform tensile strain of magnitude $2\delta_y/H$ (realized far ahead of the crack tip), multiplied by $H$.
\begin{figure}[]
\centering
\includegraphics[width = 0.75\textwidth]{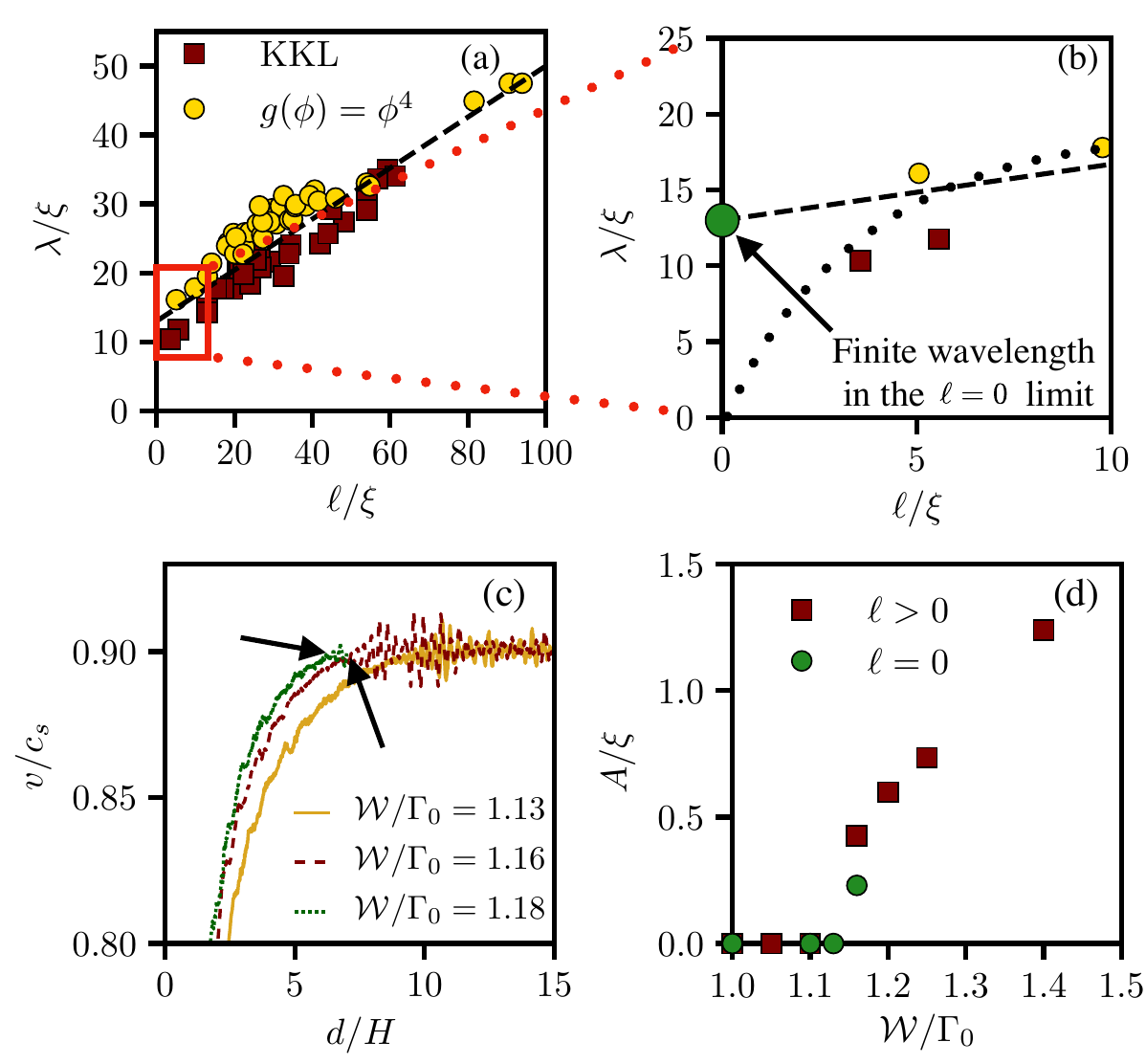}
\caption{(a) The wavelength of the oscillatory instability $\lambda$, in units of $\xi$, vs.~the dimensionless nonlinear length $\ell/\xi$, as obtained in large-scale phase-field simulations of two phase-field models. The latter differ in the choice of degradation functions, corresponding to the KKL choice, $g(\phi)\!=\!4\phi^3-3\phi^4$ and $w(\phi)=1-g(\phi)$ (brown squares) and to the modified phase-field model, where $g(\phi)\!=\!\phi^4$ and $w(\phi)\!=\!1-\phi$ (yellow circles). The nonlinear length is estimated as explained in the text. $\lambda$ varies linearly with $\ell$, as indicated by the best linear fit (dashed line), largely independently of the choice of the degradation functions and in agreement with experimental observations~\citep{Chen2017,Lubomirsky2018}. The best linear fit features a finite intercept corresponding to $\lambda\!\simeq\!13\xi$~\citep{Chen2017,Lubomirsky2018}. Some, but not all, of the data points corresponding to the KKL model overlap those reported in Fig.~2a of~\citep{Lubomirsky2018}. The red rectangle marks the region that is zoomed in on in the next panel. (b) Zooming in on the small $\ell/\xi$ regime. As explained in detail in the text, there exist two qualitatively different scenarios in relation to the $\ell/\xi\!\to\!0$ limit; one scenario (dashed line), which follows the linear fit/extrapolation of panel (a), predicts a finite intercept at $\ell\!=\!0$ (i.e.~in the absence of near tip elastic nonlinearity). The other scenario (dotted line) predicts the disappearance of the instability for $\ell\!=\!0$. Previous work failed to decide between the two qualitatively different physical scenarios. Here, using the modified phase-field model as detailed in the text and already employed in panel (a), the first scenario is supported (green circle). (c) $v/c_s$ vs.~$d/H$ for $\ell\!=\!0$ (i.e.~using the linear elastic approximation of the elastic energy functional) for different loading levels $\mathcal{W}/\Gamma_0$. For $\mathcal{W}/\Gamma_0\!=\!1.13$ no instability occurs, while for $\mathcal{W}/\Gamma_0\!=\!1.16$ (corresponding to the green circle in panel (b)) the oscillatory instability takes place (arrow) and for $\mathcal{W}/\Gamma_0\!=\!1.18$ tip-splitting occurs (arrow). (d) The normalized oscillations amplitude, $A/\xi$, vs.~the driving force ${\mathcal W}/\Gamma_0$ near the onset of instability, using the modified phase-field model for both $\ell\!=\!0$ and $\ell\!>\!0$. The results are consistent with a supercritical Hopf bifurcation, as discussed in the text. Panel (a) employs a variety of simulation box dimensions $W$ and $H$ so that the background strain remains small, and $e_{\rm c}/\mu$ is gradually varied to vary $\ell$. $W\!=\!H\!=\!200\xi$ and $e_{\rm c}/\mu\!=\!0.1$ are used elsewhere. In addition, we set $\beta\!=\!0.28$ and $\Delta\!=\!0.2\xi$.}
\label{fig:LEPF-panel}
\end{figure}

Large-scale numerical simulations of the resulting equations~\citep{Chen2017,Lubomirsky2018} revealed an oscillatory instability (cf.~Fig.~\ref{Fig:Intro-fig}a) that spontaneously initiates at a very high critical velocity $v_{\rm c}$ (cf.~Fig.~\ref{Fig:Intro-fig}f), in quantitative agreement with the experimental observations (both in terms of the existence of the instability, cf.~Fig.~\ref{Fig:Intro-fig}b, and in terms of the value of the critical velocity). Moreover, the oscillations wavelength $\lambda$ has been shown to vary linearly with the nonlinear length $\ell$~\citep{Chen2017,Lubomirsky2018} --- cf.~Fig.~\ref{fig:LEPF-panel}a ---, featuring a slope $d\lambda/d\ell$ whose value is in quantitative agreement with the corresponding experiments~\citep{Chen2017}. It is important to stress that within the theoretical-computational framework $\ell\!\sim\!\Gamma/\mu$ (at fixed $\xi$ and elastic nonlinearity) can be controllably varied by independently varying $e_{\rm c}$ (which determines the basic scale of $\Gamma$) and $\mu$, while experimentally this is far more challenging. The reason for this is that $\Gamma$ and $\mu$ may vary in a correlated manner across materials~\citep{Goldman2012}, as both depend on a basic interaction energy scale. Finally, note that the nonlinear length $\ell$ in Fig.~\ref{fig:LEPF-panel}a-b is calculated following~\citet{Lubomirsky2018}. In particular, this is done by splitting the strain energy density to its linear and nonlinear contributions, $e_{\mbox{\scriptsize{strain}}}\= e_{\mbox{\scriptsize{strain}}}^{\mbox{\tiny{\scriptsize{le}}}} + e_{\mbox{\scriptsize{strain}}}^{\mbox{\tiny{\scriptsize{nl}}}}$ and then  calculating the area that corresponds to the region where $||\partial_{\bf{F}}e_{\mbox{\scriptsize{strain}}}^{\mbox{\tiny{nl}}}/ \partial_{\bf{F}}e_{\mbox{\scriptsize{strain}}}^{\mbox{\scriptsize{le}}}||$ becomes non-negligible. Here $||\cdot||$ denotes the square root of sum of squares of all the components of the tensor. The nonlinear length is then estimated as $\ell\=\sqrt{\mathcal{A}}$, where $ \mathcal{A}$ corresponds to the area where $||\partial_{\bf{F}}e_{\mbox{\scriptsize{strain}}}/ \partial_{\bf{F}}e_{\mbox{\scriptsize{strain}}}^{\mbox{\scriptsize{le}}}||\!>\!1/2$~\citep{Lubomirsky2018}.

The extrapolation of the linear $\lambda$--$\ell$ relation to $\ell\!\to\!0$, which is inaccessible experimentally, yielded a finite intercept of $\lambda\!\simeq\!13\xi$. The very same intercept has been obtained for a different nonlinear elastic material~\citep{Lubomirsky2018}, which features a different slope $d\lambda/d\ell$ compared to brittle neo-Hookean materials. If this extrapolation is physically valid, it has dramatic implications; it suggests that the oscillatory instability exists also in the absence of elastic nonlinearity ($\ell\!\to\!0$), in which case the oscillations wavelength is inherited from the other intrinsic length scale in the problem, i.e.~the dissipation length $\xi$. The flexibility of the theoretical-computational phase-field framework allows one to probe the $\ell\!\to\!0$ limit, going significantly beyond experiments. The ratio $\ell/\xi$ can be reduced down to $O(10^{-2})$ by reducing $e_{\rm c}$, as shown in~Fig.~\ref{fig:LEPF-panel}b, and the wavelength $\lambda$ seems to follow the linear dependence on $\ell$, in agreement with the prediction based on the linear approximation. To decisively resolve the $\ell\!\to\!0$ limit, one should actually set $\ell\=0$. This cannot be achieved through $e_{\rm c}$, which has to remain finite, but rather by controlling the prefactor in the relation $\ell\!\sim\!\Gamma/\mu$, which depends on nonlinear elastic coefficients.

Since the prefactor in the relation $\ell\!\sim\!\Gamma/\mu$ vanishes identically in the absence of elastic nonlinearity, $\ell\=0$ can be achieved by setting all nonlinear elastic coefficients to zero, i.e.~by a priori using the linear elastic approximation of the nonlinear elastic functional $e_{\mbox{\scriptsize{strain}}}$. For 2D brittle neo-Hookean materials described by Eq.~\eqref{eq:neoHookean}, the linear elastic approximation takes the form $e^{\mbox{\tiny{le}}}_{\mbox{\scriptsize{strain}}}\=\mu\,\bar{e}^{\mbox{\tiny{le}}}_{\mbox{\scriptsize{strain}}}\=\mu([\tr ({\bm \epsilon})]^2 + \tr({\bm\epsilon^2}))$, where ${\bm \epsilon}$ is the linear elastic strain tensor whose components are $\epsilon_{ij}\=\tfrac{1}{2}(\partial_j{u_i} + \partial_i{u_j})$. Using $e^{\mbox{\tiny{le}}}_{\mbox{\scriptsize{strain}}}\=\mu\,\bar{e}^{\mbox{\tiny{le}}}_{\mbox{\scriptsize{strain}}}$ inside Eq.~\eqref{Eq:Lagrangian} corresponds to a material with $\ell\=0$; performing such calculations did not yield an oscillatory instability. This result suggests a qualitatively different physical scenario in which the oscillatory instability disappears in the absence of elastic nonlinearity $\ell\!\to\!0$, cf.~the dotted line in~Fig.~\ref{fig:LEPF-panel}b, compared to the linear extrapolation scenario in which the instability exists in this limit, cf.~the dashed line in~Fig.~\ref{fig:LEPF-panel}b, and is controlled by the dissipation length $\xi$.

How can one decide between these two mutually exclusive and qualitatively different physical scenarios? As the equations of motion are numerically resolved on a square lattice/grid, we cannot exclude the possibility that the oscillatory instability is spuriously suppressed for $\ell\=0$, where the oscillations amplitude is expected to be small, due to lattice pinning (see discussion in Sec.~\ref{sec:intro} and Sec.~\ref{sec:PF}). That is, it is conceivable that the crack is trapped at a numerical lattice plane and hence cannot oscillate when its oscillations amplitude is vanishingly small. To address this possibility, and inspired by~\citet{Dolbow2019}, we formulate a modified phase-field model that employs $g(\phi)\=\phi^4$ and $w(\phi)\=1-\phi$. This choice of degradation functions is expected to increase the size of the effective cohesive zone compared to the KKL choice, which in turn is expected to reduce lattice pinning effects. Additional and more detailed discussion of this modified phase-field formulation will be presented elsewhere~\citep{Vasudevan2020}.

For a finite nonlinear length $\ell$, it is known that the oscillatory instability is controlled by $\ell$, where the cohesive zone and its characteristic scale $\xi$ play a secondary role. Consequently, we expect the modified phase-field formulation to give rise to the very same oscillatory instability discussed above in relation to the KKL degradation functions. To test this, we performed simulations with the modified phase-field formulation for $\ell\!>\!0$, i.e.~with the nonlinear elastic energy functional of Eq.~\eqref{eq:neoHookean}, and superimposed in Fig.~\ref{fig:LEPF-panel}a the oscillations wavelength $\lambda$ vs.~$\ell$ for this model (yellow circles) on top of the corresponding results for the KKL model (brown squares). The two data sets nearly collapse, clearly demonstrating the expected independence of $\lambda(\ell)$ on the details of the dissipation/cohesive zone when $\ell\!>\!0$. The main merit of the modified model, in the present context, would be to explore its behavior for $\ell\=0$; if indeed the modified degradation functions reduce lattice pinning effects, we expect this model to distinguish between the two qualitatively different physical scenarios discussed above, i.e.~to unambiguously show whether the oscillatory instability exists in the $\ell\!\to\!0$ limit (the dashed line hypothesis in Fig.~\ref{fig:LEPF-panel}b) or disappears (the dotted line hypothesis in Fig.~\ref{fig:LEPF-panel}b). Performing this calculation clearly reveals an oscillatory instability for $\ell\=0$, with a wavelength $\lambda\!\approx\!13\xi$ (marked by the large green circle in Fig.~\ref{fig:LEPF-panel}b), in quantitative agreement with the linear extrapolation prediction (the dashed line in Fig.~\ref{fig:LEPF-panel}b, which intercepts the $\ell\=0$ line exactly at this value). Furthermore, in Fig.~\ref{fig:LEPF-panel}c we present $v/c_s$ vs.~$d/H$ for $\ell\=0$, showing that the oscillatory instability emerges as a critical velocity $v_c$ is reached (with increasing loading level ${\cal W}/\Gamma_0$), in full agreement with the $\ell\!>\!0$ results of Fig.~\ref{Fig:Intro-fig}f. We also show that further increasing ${\cal W}/\Gamma_0$ leads to tip-splitting, yet again in agreement with the $\ell\!>\!0$ results of Fig.~\ref{Fig:Intro-fig}f.

The results presented in Fig.~\ref{fig:LEPF-panel}b-c have far-reaching implications. Most notably, they show that the oscillatory instability crucially depends on the existence of an intrinsic length scale associated with the breakdown of linear elasticity near crack tips, independently of whether it is related to nonlinear elasticity or to dissipation. In particular, they show that near tip elastic nonlinearity is not a necessary condition for the existence of instability, which is expected to be observed also in very stiff materials in 2D. Furthermore, the fact that the wavelength in the $\ell\=0$ limit, $\lambda\!\simeq\!13\xi$, is significantly larger than the bare dissipation length $\xi$ reflects the strongly inertial nature of the instability, where elastodynamic effects emerging for $v/c_s\!\sim\!O(1)$ renormalize the magnitude of the region in which LEFM breaks down ahead of the tip. This effect is shown in~\ref{sec:AppendixC} to be consistent with direct observations of the near tip fields of straight $\ell\=0$ cracks propagating at high velocities.

The resolution of the $\ell\!\to\!0$ limit also allows to determine the nature of the bifurcation occurring at the onset of instability, which remained previously unknown. To address this question, we set $\ell\=0$ together with $g(\phi)\= \phi^4$ and $w(\phi)\=1-\phi$ (as in the discussion above), and calculated the amplitude $A$ of the oscillations close to the onset of instability as a function of the dimensionless driving force for fracture ${\cal W}/\Gamma_0$. For relatively low driving forces, the crack does not reach the critical oscillations velocity $v_{\rm c}$, i.e.~$A\=0$. With increasing ${\cal W}/\Gamma_0$, the oscillations emerge and their amplitude can be extracted. The results are presented in~Fig.~\ref{fig:LEPF-panel}d, where it is shown that $A$ increases from zero at the onset of instability in a continuous manner for both $\ell\=0$ and $\ell\!>\!0$, but apparently with a discontinuous derivative. This behavior is characteristic of a supercritical Hopf bifurcation~\citep{strogatz2018nonlinear}, in line with the theoretical predictions of~\citet{Bouchbinder.2009a}. Furthermore, the vanishingly small oscillations amplitude at onset indeed supports the idea that the KKL model did not feature an oscillatory instability for $\ell\=0$ due to numerical lattice pinning.

The established properties of the oscillatory instability and its theoretical understanding, most notably its dependence on intrinsic material length scales, clearly suggest that the salient features of the instability are independent of the fracture energy $\Gamma(v)$. In the presence of near tip nonlinear elasticity, the major predicted effect of $\Gamma(v)$ is a renormalization of the oscillations wavelength by $\Gamma(v_{\rm c})$, according to $\lambda\!-\!13\xi\!\sim\!\ell\!\sim\!\Gamma(v_{\rm c})/\mu$. Moreover, the slope $d\lambda/d\ell$ is predicted to be independent of the functional form of $\Gamma(v)$. These predictions can be readily tested within the phase-field approach as $\Gamma(v)$ can be varied following Eq.~\eqref{eq:v-Gamma}, by either varying the parameter $\beta\!\sim\!\chi^{-1}$ or by varying the degradation functions that affect the integral on the right-hand-side of Eq.~\eqref{eq:v-Gamma}.

\begin{figure}[h]
\includegraphics[height =2.5 in]{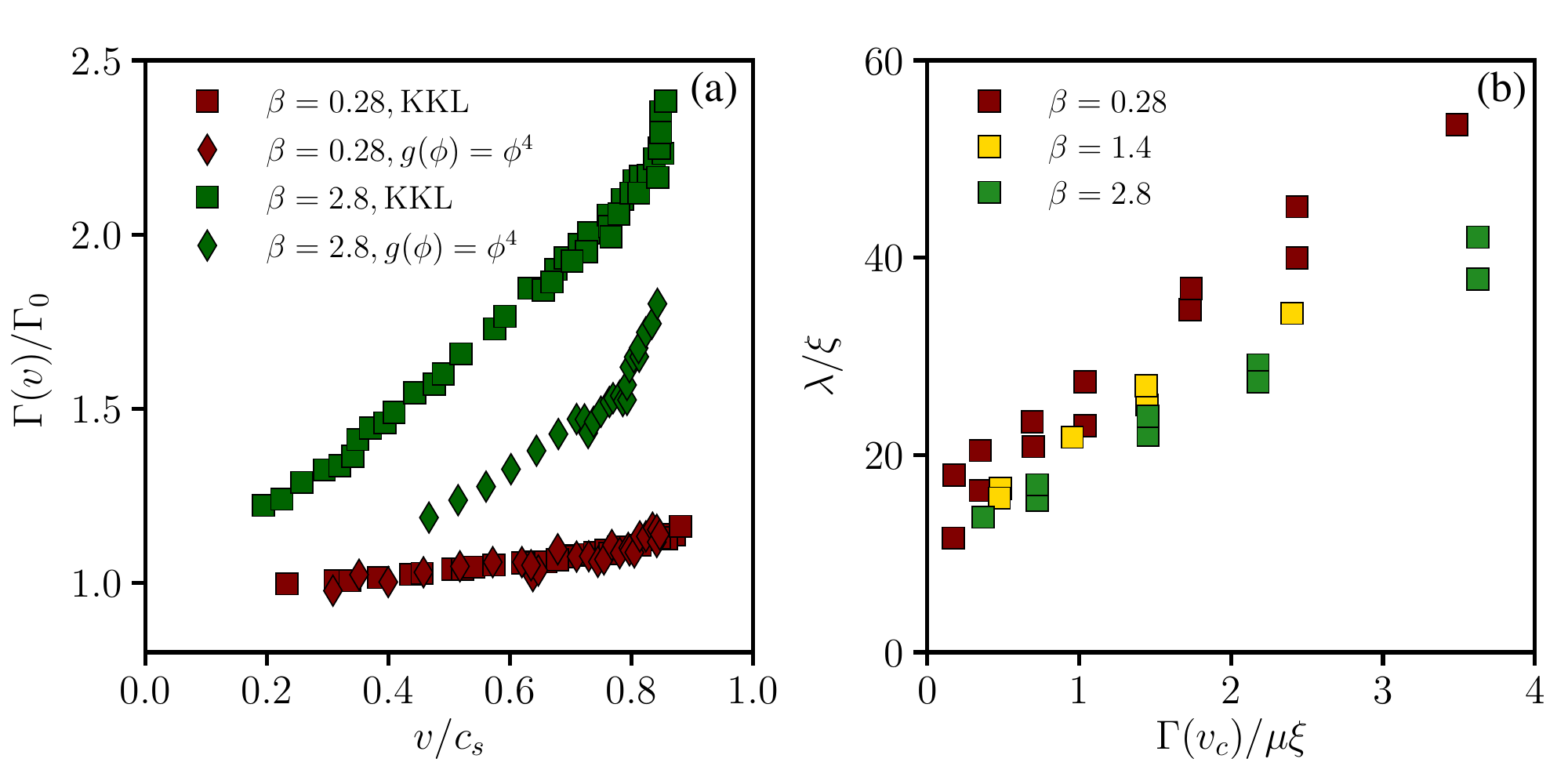}
\centering
\caption{(a) The normalized fracture energy $\Gamma(v)/\Gamma_0$ vs.~$v/c_s$ for two values of $\beta$, which are separated by an order of magnitude (see legend) and for two choices of degradation functions, $g(\phi)\!=\!4\phi^3\!-\!3\phi^4$ and $w(\phi)\!=\!1\!-\!g(\phi)$ (termed KKL) and the modified one, i.e.~$g(\phi)\!=\!\phi^4$ and $w(\phi)\!=\!1\!-\!\phi$ (see text for additional details). (b) The normalized oscillations wavelength $\lambda/\xi$ vs.~$\Gamma(v_{\rm c})/\mu\xi$ (which is proportional to $\ell/\xi$) for the KKL choice and various values of $\beta$ (see legend). Compare the results to those of Fig.~\ref{fig:LEPF-panel}a and see text for discussion. The simulation box features $H\!=\!300\xi$ and $W\!=\!900\xi$, and the spatial discretization size is $\Delta\!=\!0.2\xi$, for both panels (a) and (b). In panel (a) the linear elastic strain energy density $e^{\mbox{\tiny{le}}}_{\mbox{\scriptsize{strain}}}$ corresponding to Eq.~\eqref{eq:neoHookean} (see text for additional details) is used with $e_{\rm c}/\mu\!=\!0.5$, while in panel (b) $e_{\mbox{\scriptsize{strain}}}$ of Eq.~\eqref{eq:neoHookean} is used and $e_{\rm c}/\mu$ is gradually varied in order to vary $\Gamma(v_{\rm c})/\mu\xi$.}
\label{Fig:Gamma-v}
\end{figure}

In Fig.~\ref{Fig:Gamma-v}a we present $\Gamma(v)$ corresponding to a $10$-fold variation in $\beta$, and to the two choices of the degradation functions $g(\phi)$ and $w(\phi)$ discussed above (the KKL one, and the modified one, i.e.~$g(\phi)\=\phi^4$ and $w(\phi)\=1-\phi$). $\Gamma(v)$ was calculated through the relation $\Gamma(v)\={\cal J}(v)/v$, where the J-integral ${\cal J}(v)$ is given by ${\cal J}(v)\=\int_{C}[(U+T)\,v\,n_x+P_{ij}\,\partial_t{u_i}\,n_j]dC$~\citep{Bouchbinder.09, Livne2010, Freund.98, Nakamura1985}. Here $C$ is a close contour surrounding the crack tip outside of the dissipation zone (i.e.~predominantly along a $\phi(x,y)\=1$ path), ${\bm n}\= (n_x,n_y)$ is the outward normal to the contour and $P_{ij}$ are the components of the first Piola-Kirchhoff stress tensor, cf.~Eq.~\eqref{Eq:P_def_Appendix} (see details about the numerical implementation of the J-integral in~\ref{sec:AppendixB}). It is observed that $\Gamma(v)$ varies quite significantly, from being nearly flat for the smallest $\beta$ value used to exhibiting substantial $v$-dependence for the largest one. Moreover, note that for small $\beta$ (here $\beta\= 0.28$) $\Gamma(v)$ is insensitive to the choice of degradation functions and is almost independent of $v$. For larger values of $\beta$ (here $\beta\=2.8$), not only $\Gamma(v)$ exhibits rather strong $v$-dependence, but it also depends on the degradation functions that control the phase-field behavior in the cohesive zone (compare the green squares and diamonds in Fig.~\ref{Fig:Gamma-v}a, corresponding to the two choices of the degradation functions).

As the largest variation of $\Gamma(v)$ with $\beta$ is observed for the KKL degradation functions (squares in Fig.~\ref{Fig:Gamma-v}a), we focused on this case and performed extensive calculations for three values of $\beta$, spanning an order of magnitude (see legend of Fig.~\ref{Fig:Gamma-v}b), using Eq.~\eqref{eq:neoHookean}. For each calculation, we accelerated the crack to the critical velocity $v_{\rm c}$ for the onset of oscillations and extracted the oscillation wavelength $\lambda$. In Fig.~\ref{Fig:Gamma-v}b we plot $\lambda/\xi$ vs.~$\Gamma(v_{\rm c})/\mu\xi$ for the three values of $\beta$ indicated in the legend. It is observed that despite the large variation in $\Gamma(v)$ (cf.~Fig.~\ref{Fig:Gamma-v}a), the relation $\lambda\!-\!13\xi\!\sim\!\ell\!\sim\!\Gamma(v_{\rm c})/\mu$ is satisfied to a fairly good degree independently of $\Gamma(v)$, with a slope $d\lambda/d\ell$ (which depends on the form of near tip elastic nonlinearity, kept fixed in these calculations) that is also independent of it, as predicted theoretically. These results, together with the existence of the oscillatory instability in the $\ell\!\to\!0$ limit, substantiate the universal nature of the oscillatory instability, which is expected to be observed in any material in 2D.

\section{The ultra-high velocity tip-splitting instability: Relations to the wave-speed inside the dissipation zone}
\label{sec:tip-splitting}

\begin{figure}[hbt!]
\includegraphics[width=\textwidth]{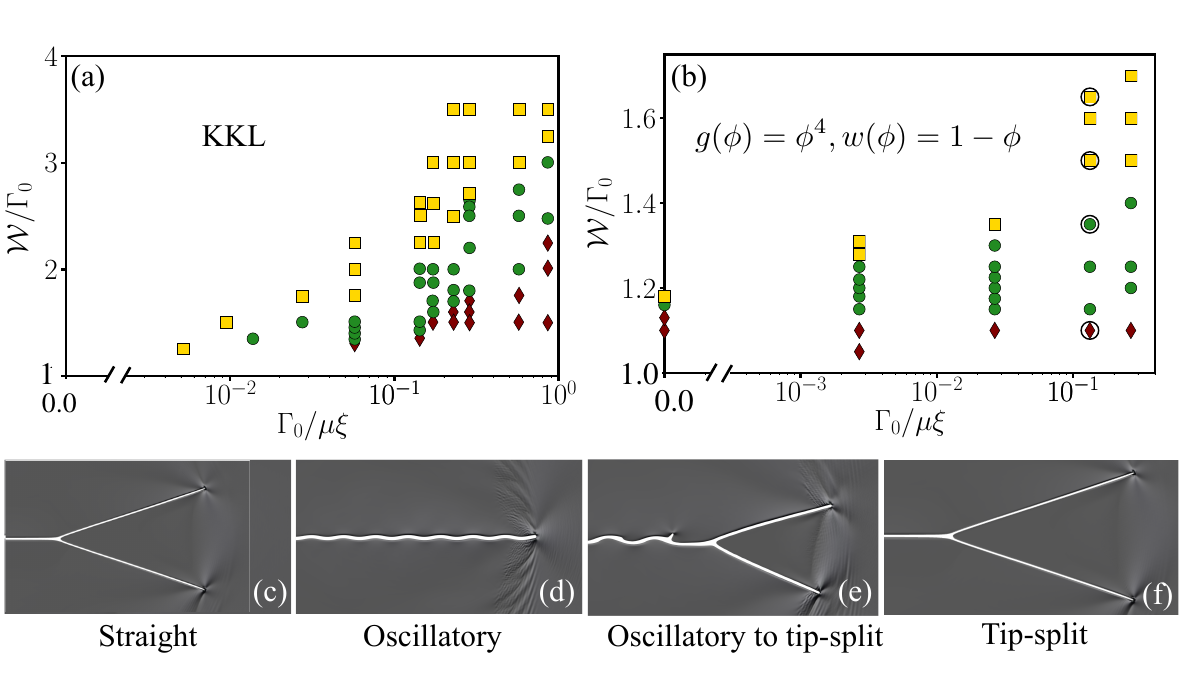}
\centering
\caption{(a) The phase diagram of 2D dynamic fracture, in the ${\cal{W}}/\Gamma_0$---$\Gamma_0/\mu\xi$ plane, for materials featuring near tip nonlinear neo-Hookean elasticity, and using the KKL degradation functions (recall that $\ell/\xi\!\sim\!\Gamma_0/\mu\xi$ and compare to~\citet{Lubomirsky2018}). As $\Gamma_0$ is the fracture energy at the onset of crack propagation, we focus on ${\cal W}/\Gamma_0\!>\!1$. For a fixed $\ell/\xi\!\sim\!\Gamma_0/\mu\xi$, straight crack states exist at small driving forces ${\cal W}/\Gamma_0$ (diamonds), oscillatory crack states (circles) exist for larger ${\cal W}/\Gamma_0$ (required to surpass the critical oscillations velocity $v_{\rm c}$) and tip-split crack states, which include both the oscillatory to tip-split states and straight to tip-split states, exist at yet higher driving forces (squares). The range of driving forces ${\cal W}/\Gamma_0$ for which straight and oscillatory cracks exist diminishes with decreasing $\Gamma_0/\mu\xi$, as discussed in~\citet{Lubomirsky2018}, where it was also shown that the presented topology of the phase diagram is independent of the form of near tip elastic nonlinearity. See additional discussion of the phase diagram in the text. (b) The same as in panel (a), but with the modified choice of degradation functions $g(\phi)\!=\!\phi^4$ and $w(\phi)\!=\!1-\phi$. The topology of the phase diagram remains the same as in panel (a), but due to reduced lattice pinning, we can obtain the phase diagram also for $\ell/\xi\sim\Gamma_0/\mu\xi\!\to\!0$ and even exactly at $\ell\!=\!0$. (c)-(f) Snapshots of the crack states (shown in the undeformed coordinates) representative of each distinct region in the phase diagram, corresponding to straight (panel (c)), oscillatory (panel (d)), oscillatory to tip-split (panel (e)) and straight to tip-split (panel (f)) crack states. These snapshots correspond to the symbols with black edge markers shown in panel (b). Note the frustrated tip-splitting event in panel (e), taking place prior to the actual tip-splitting. In all of the simulations reported here we used $\beta\!=\!0.28$ and $\Delta\!=\!0.2\xi$. A simulation box of $W\!=\!H\!=\!200\xi$ is used for small $\Gamma_0/\mu\xi$ values, which is gradually increased to $W\!=\!H\!=\!600\xi$ for larger $\Gamma_0/\mu\xi$, in order to maintain a small background strain.}
\label{Fig:phase-diagram}
\end{figure}

As discussed above in relation to Figs.~\ref{Fig:Intro-fig}c,f and~\ref{fig:LEPF-panel}c, upon increasing the driving force ${\cal W}/\Gamma_0$ for fracture, cracks are predicted to accelerate faster and to yet higher velocities, and feature a tip-splitting instability, either after the onset of oscillations or even prior to it. This behavior is supported by experiments, cf.~Fig.~\ref{Fig:Intro-fig}d. The observation of tip-split crack states, together with the previously discussed oscillatory crack states, allow one to construct a comprehensive phase diagram for 2D dynamic fracture, which is presented in Figs.~\ref{Fig:phase-diagram}a-b for both the KKL and modified ($g(\phi)\=\phi^4$, $w(\phi)\= 1-\phi$) choices of the degradation functions, respectively. These phase diagrams highlight the different crack states attained as a function of the intrinsic length scale ratio $\ell/\xi\!\sim\!\Gamma_0/\mu\xi$ and the normalized driving force ${\cal W}/\Gamma_0$. These include straight crack states (diamonds), oscillatory crack states (circles) and oscillatory/straight cracks followed by tip-splitting (squares). Snapshots of the different crack states (in the undeformed coordinates) are shown in Figs.~\ref{Fig:phase-diagram}c-f. Note that the topology of the phase diagram is independent of the choice of the degradation functions, though some quantitative differences are evident. Most notably, since $\Gamma(v)$ is significantly smaller for the modified model (with $g(\phi)\=\phi^4$ and $w(\phi)\=1-\phi$, cf.~Fig.~\ref{Fig:Gamma-v}a), lower driving force levels ${\cal W}/\Gamma_0$ are needed to reach the critical velocity for these instabilities, and hence in this case the $y$-axis range is smaller (compare Figs.~\ref{Fig:phase-diagram}a-b). Finally, note that the phase diagram in Fig.~\ref{Fig:phase-diagram}b extends all the way to the $\ell\=0$ limit (by overcoming lattice pinning, as discussed above), featuring the same sequence of transitions as for $\ell\!>\!0$.

One feature of the tip-splitting instability is the angle formed by the two branches. We define the tip-splitting angle as half of the angle between the two symmetric branches, e.g.~those shown in Fig.~\ref{Fig:phase-diagram}f. In~\citet{Katzav2007theory}, using the Griffith energy criterion and the principle of local symmetry in the framework of LEFM, a tip-splitting angle of $27^\circ$ that is independent of the critical tip-splitting velocity has been predicted. To make contact with this prediction --- despite the fact that the bound on the critical tip-splitting velocity predicted in~\citet{Katzav2007theory}, of about half $c_s$, is substantially smaller than our observed one (around $0.9c_s$) ---, we measured the tip-splitting angle in our simulations over a length sufficiently larger than both $\xi$ and $\ell$ (for LEFM to be relevant/applicable), and sufficiently smaller than the system height $H$ (to avoid curving of the branches due to interactions with the boundaries). We find tip-splitting angles in the range $24^\circ\pm 2^\circ$, weakly dependent on material and loading parameters. This result appears to be in reasonable agreement with the prediction of~\citet{Katzav2007theory}, though we stress again that the same LEFM considerations seem to seriously fail to predict the tip-splitting critical velocity.

Our next goal in this section is to gain physical insight into the origin of this ultra-high velocity instability, which together with the discussion of the oscillatory instability in Sec.~\ref{sec:oscillatory}, would offer a comprehensive understanding of dynamic instabilities in 2D fracture. To set the stage for this discussion, let us recall the form of the kinetic energy contribution to the Lagrangian, $T\=\int\!\tfrac{1}{2}\,g(\phi)\mu\,[c_{\rm pz}(\phi)]^{-2}(\partial{\bm u}/\partial t)^2 dV$, where $c_{\rm{pz}}(\phi)$ is given in Eq.~\eqref{Eq:waveSpeedxi}. In the analysis above, as well as in~\citet{Chen2017,Lubomirsky2018}, we set $f(\phi)\=g(\phi)$ that implies $c_{\rm{pz}}(\phi)\=c_s$, i.e.~the wave-speed inside the dissipation zone equals its elastic bulk value. It was argued that this situation is representative of realistic dissipative processes, e.g.~plastic deformation, which generally do not lead to a significant softening of elastic moduli (in polycrystals, known to undergo strain hardening, elastic moduli actually stiffen). It is precisely this choice that allowed cracks in this framework to accelerate to unprecedentedly high velocities, which in turn allowed the oscillatory instability to be elucidated.

In previous work~\citep{karma2001phase,Karma2004,Henry.08}, the relation $f(\phi)\=1$ has been employed, which implies that $c_{\rm{pz}}(\phi)$ in Eq.~\eqref{Eq:waveSpeedxi} degrades together with the effective modulus $g(\phi)\mu$. In such 2D models, cracks are known to undergo a tip-splitting instability at moderate velocities in the range of $0.4c_s\!-\!0.5c_s$, in sharp contrast to 2D experiments in which cracks accelerate to much higher velocities until oscillations set in. These observations suggest that material inertia, which plays a central role in the conversion of elastic strain energy into fracture energy at high propagation velocities, is the limiting factor that controls the 2D tip-splitting instability. In other words, we suggest that tip-splitting occurs when the crack velocity $v$ approaches the characteristic wave-speed inside the dissipation zone. To see this more formally, we express the kinetic energy $T$ in a co-moving frame of reference of a crack propagating steady at a velocity $v$ along the $x$-direction, i.e.~$T\=\int\!\tfrac{1}{2}\,g(\phi)\mu\,[v/c_{\rm pz}(\phi)]^2(\partial{\bm u}/\partial x)^2 dV$. This expression suggests that the model's behavior depends on $v/c_{\rm pz}$, where $c_{\rm pz}$ is a characteristic value of $c_{\rm pz}(\phi)$ inside the dissipation zone, and consequently that the tip-splitting is affected by $c_{\rm pz}$.

\begin{figure}[t!]
\includegraphics[height=4in]{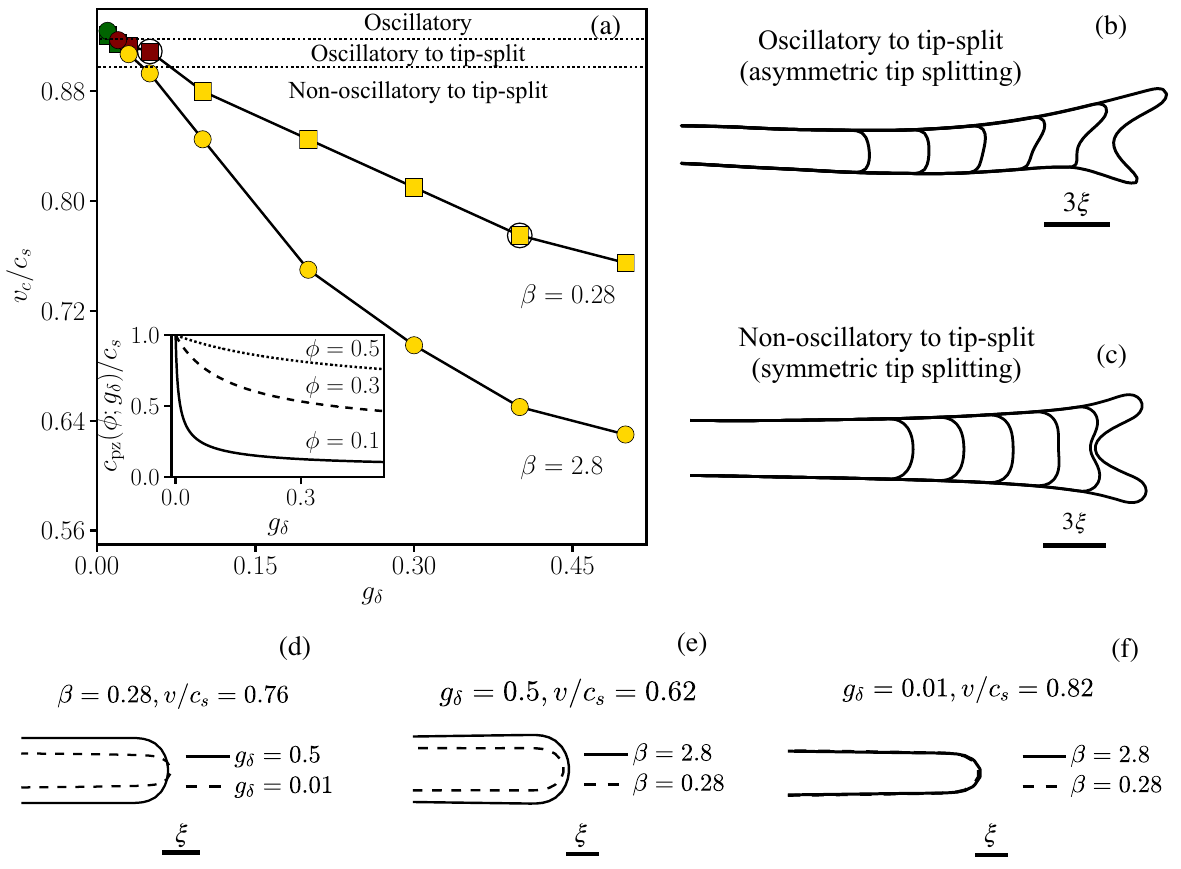}
\centering
\caption{(a) The normalized critical velocity $v_{\rm c}/c_s$ at which a straight crack loses stability is plotted against $g_\delta$ (cf.~Eq.~\eqref{eq:g_delta}), for two different values of $\beta$.
Yellow symbols correspond to a tip-splitting instability, which emerges directly from a straight crack state, while brown symbols correspond to tip-splitting that emerges from an oscillatory crack state and green symbols correspond to the oscillatory instability. Inset: $c_{\rm pz}(\phi; g_\delta)/c_s$, defined through Eqs.~\eqref{Eq:waveSpeedxi} and~\eqref{eq:g_delta}, as a function of $g_\delta$ for three values of $\phi\!<\!1$ (as indicated on the figure). (b) $\phi\!=\!1/2$ contours, plotted at equal time intervals, for a crack that asymmetrically tip-splits after the onset of oscillations (corresponding to the encircled brown square in panel (a)). (c) The same as panel (b), but for a crack that symmetrically tip-splits directly from a straight crack state (corresponding to the encircled yellow square in panel (a)). (d) $\phi\!=\!1/2$ contours for two different values of $g_\delta$, and fixed $\beta$ and $v/c_s$ (see figure for the values and the text for discussion). (e) The same as panel (d), but for two values of $\beta$, and fixed $g_\delta\!=\!0.5$ and $v/c_s$ (see figure for the values and the text for discussion). (f) The same as panel (e), but for $g_\delta\!=\!0.01$. The simulation box for panels (a)-(f) features $W\!=\!H\!=\!300\xi$, using the KKL degradation functions. In panel (a), $\Gamma_0/\mu\xi\!=\!0.287$ is used, and for $\beta\!=\!0.28$ the driving force is fixed at ${\cal W}/\Gamma_0\!=\!2.0$, while for $\beta\!=\!2.8$ we used ${\cal W}/\Gamma_0\!=\!3.8$. Panels (b)-(f) are plotted using the results shown in panel (a). A grid spacing of $\Delta\!=\!0.2\xi$ is used for all simulations and scale bars in units of $\xi$ are added.}
\label{Fig:f-phi}
\end{figure}

To test this idea we introduce a control parameter $g_\delta$ that allows to continuously extrapolate between the $f(\phi)\=g(\phi)$ and $f(\phi)\=1$ limits. This is done by defining
\begin{equation}
f(\phi; g_\delta) = \frac{ g(\phi) + g_\delta}{1 + g_\delta} \ ,
\label{eq:g_delta}
\end{equation}
where $f(\phi; g_\delta\=0)\=g(\phi)$ and $f(\phi; g_\delta\!\gg\!1)\!\to\!1$. Consequently, we define $c_{\rm pz}(\phi; g_\delta)/c_s\!\equiv\!\sqrt{g(\phi)/f(\phi; g_\delta)}$, which is plotted in the inset of Fig.~\ref{Fig:f-phi}a for the KKL choice of degradation functions. We performed calculations for neo-Hookean materials for a wide range of $g_\delta$ values, as shown in Fig.~\ref{Fig:f-phi}a for $\beta\=0.28$ and $\beta\=2.8$, where $v_{\rm c}/c_s$ is plotted against $g_\delta$ ($v_c$ is the critical velocity for an instability, independently of its nature, i.e.~whether it corresponds to oscillations or tip-splitting). For each $\beta$ value, the driving force was fixed, where ${\cal W}/\Gamma_0\=2.0$ (cf.~Fig.~\ref{Fig:Intro-fig}) for $\beta\=0.28$ and ${\cal W}/\Gamma_0\=3.8$ for $\beta\=2.8$ were used. The values of ${\cal W}/\Gamma_0$ are chosen such that the oscillatory instability emerges in the $g_{\delta}\!\to\!0$ limit, as used previously throughout the paper.

For very small values of $g_\delta$, i.e.~in the $g_\delta\!\to\!0$ limit, the results reported on in Fig.~\ref{Fig:Intro-fig}a are reproduced; that is, the crack accelerates to very high velocities and exhibits an oscillatory instability (green symbols). With increasing values of $g_{\delta}$, a tip-splitting instability is observed at smaller $v_{\rm c}$, either preceded by oscillations (brown symbols) or emerging directly from a straight crack (yellow symbol), where the latter occurs for sufficiently large $g_\delta$. This sequence of transitions with increasing $g_\delta$ lends support to the role played by $c_{\rm pz}$ in determining the crack velocity needed for tip-splitting. For (relatively) large $g_\delta$, where $c_{\rm pz}$ is small, the critical tip-splitting velocity is small and tip-splitting is observed as the crack accelerates (yellow symbols). As $g_\delta$ decreases, the critical tip-splitting velocity increases, until at some point it becomes larger than the critical oscillations velocity and the latter is observed (green symbols).

The trend of a decreasing tip-splitting velocity $v_{\rm c}$ with increasing $g_\delta$ (i.e.~decreasing wave-speed $c_{\rm{pz}}$) is observed for both low ($\beta\=0.28$) and high ($\beta\=2.8$) fracture energies, with a larger $\beta$ yielding a lower tip-splitting velocity for the same $g_\delta$ (cf.~Fig.~\ref{Fig:f-phi}a). To shed light on the mechanism of the tip-splitting instability, we show in Figs.~\ref{Fig:f-phi}b-c a sequence of  $\phi\=1/2$ contours at equal time intervals just before tip-splitting. In Fig.~\ref{Fig:f-phi}b, tip-splitting occurs asymmetrically, as it is preceded by an oscillatory behavior that breaks the reflection symmetry relative to the straight crack propagation axis, while in Fig.~\ref{Fig:f-phi}c tip-splitting occurs symmetrically, directly from a non-oscillatory straight crack. Figures~\ref{Fig:f-phi}b-c correspond to the two encircled symbols in panel (a). For both of these cases, as the crack approaches the threshold velocity $v_{\rm c}$ for tip-splitting, the crack tip blunts, suggesting a picture in which tip-splitting can be seen as an exacerbated form of tip-blunting. Tip-blunting, in turn, is expected to be more prominent as $c_{\rm{pz}}$ decreases, simply because the radiating energy away from the tip becomes more limited, leading to an increase in the amount of fracture surfaces generated (which is larger for blunter tips).

This picture is further tested in Fig.~\ref{Fig:f-phi}d, showing the crack tip shape ($\phi\=1/2$ contours) for $g_\delta\=0.01$ and $g_\delta\=0.5$, for fixed values of $\beta$ and $v/c_s$. It is observed that indeed reducing $g_\delta$, which increases $c_{\rm{pz}}$, is accompanied by reduced tip-blunting. A sharper crack-tip, in turn, suppresses tip-splitting in the $g_{\delta}\!\to\!0$ limit and enables the crack to reach ultra-high velocities that exceed the threshold for the oscillatory instability (green symbols in Fig.~\ref{Fig:f-phi}a). The crack tip shape is also influenced by energy dissipation at the crack tip. In particular, increasing the rate of dissipation (through $\beta$) increases the size of the process/cohesive zone, accompanied by a blunter crack tip. This is observed in Fig.~\ref{Fig:f-phi}e that compares the crack tip shapes for $g_\delta\= 0.5$ and two different values of $\beta$, an order of magnitude apart. The tip shape corresponding to the larger $\beta$ value is significantly blunter. As explained, a crack tip that is blunter is expected to tip-split at a lower critical velocity, which is clearly observed in Fig.~\ref{Fig:f-phi}a, where for larger $g_\delta$, $v_{\rm c}$ is smaller for the larger $\beta$. This trend is similar to the one previously reported on in mode-III dynamic fracture simulations~\citep{Karma2004}, where an increase rate of dissipation was found to promote tip-splitting in the $ g_\delta\!\to\!\infty$ limit. Note, however, that in the $g_\delta\!\to\!0$ limit, $\beta$ has a negligible effect and the crack tip shape is almost independent of it, as shown in Fig.~\ref{Fig:f-phi}f.

\section{Discussion and concluding remarks}
\label{sec:summary}

In this paper, we used phase-field simulations to investigate the role of intrinsic material length and time scales on the emergence of oscillatory and tip-splitting instabilities in 2D dynamic fracture.
The two basic length scales, which are absent in LEFM, include the scale $\xi$ of the dissipation zone where elastic energy is transformed irreversibly into new fracture surfaces and a nonlinear length $\ell$ that is a measure of the distance from the crack tip at which elastic nonlinearity becomes significant and modifies the $1/\sqrt{r}$ divergence of the linear-elastic fields. The basic time scale $\tau$, which is only indirectly present in LEFM through the dependence of the fracture energy on crack velocity, $\Gamma(v)$, controls the rate of energy dissipation inside the process zone. This time scale is only physically meaningful when compared to the characteristic time $\sim\!\xi/c_s$ for elastic waves to traverse the dissipation zone $\sim\!\xi/c_s$. When $\tau\!\ll\!\xi/c_s$ ($\beta\=\tau\,c_s/\xi\!\ll\!1$), dissipation rate has a negligible effect on the crack dynamics and $\Gamma(v)$ is nearly independent of $v$, while in the opposite limit $\beta\!\gg\!1$, dissipation is sluggish and becomes rate limiting, thereby causing $\Gamma(v)$ to increase with $v$.
Our simulations exploited a recently developed Lagrangian phase-field formulation~\citep{Chen2017} that incorporates a degradation function in the kinetic energy so as to maintain the wave-speeds constant inside the dissipation zone, thereby enabling cracks to accelerate without tip-splitting to the range of ultra-high speed approaching $c_s$, where oscillations are observed experimentally in thin brittle materials~\citep{Livne.07}, and reproduced remarkably by phase-field simulations in the same velocity range~\citep{Chen2017,Lubomirsky2018}. Simulations also reproduced an experimentally observed tip-splitting behavior that causes a new crack to emerge asymmetrically on one side of a propagating oscillatory crack~\citep{Lubomirsky2018}.

The present results shed additional light on both the oscillatory and tip-splitting instabilities. First, the results further support the fundamental role of elastic nonlinearity in the genesis of the oscillatory instability with an intrinsic (i.e.~system-size-independent) wavelength, by showing that this instability occurs even in the limit $\ell\!\ll\!\xi$ where linear elasticity holds outside of the dissipation zone. This limit was investigated here by simulating rapid fracture in a purely linear-elastic phase-field formulation, where nonlinearity is only present inside the dissipation zone where material deformation and the phase-field are coupled. Importantly, the oscillatory wavelength in this model is an order of magnitude larger than $\xi$ and coincides with the extrapolation to the $\ell/\xi\!\rightarrow\!0$ limit of phase-field simulations with nonlinear neo-Hookean elasticity, in strong support of our interpretation that this coupling acts as an effective form of nonlinearity, on equal footing with neo-Hookean or Saint Venant-Kirchhoff nonlinear elasticity~\citep{Lubomirsky2018}. This effect is explicitly addressed in~\ref{sec:AppendixC}. Interestingly, therein we provide evidence that the lengthscale that determines the wavelength of the oscillatory instability in the $\ell\=0$ limit --- which is an order to magnitude larger than $\xi$ --- appears to be comparable to the size of the region around the crack tip where the extensional strain exhibits a non-monotonous behavior (cf.~Fig.~\ref{Fig:tip_fields}) in this very same $\ell\=0$ limit (where the LEFM extensional strain is known to become negative, see discussion in~\ref{sec:AppendixC}).

On a more technical side, one limitation of the present study is that we were only able to demonstrate the existence of an oscillatory instability in a purely linear-elastic phase-field model for a particular choice of degradation functions in the potential energy that reduces lattice trapping of cracks along lattice planes. This was accomplished by choosing a combination of functions $g(\phi)$ and $w(\phi)$ that increases the length of the dissipation zone along the crack propagation direction, thereby pushing the discontinuity of material displacement on the numerical lattice/grid scale further behind the crack tip. We expect, however, the same result to hold true in other formulations or numerical implementations on unstructured grids that sufficiently reduce lattice pinning to permit small amplitude oscillations to be numerically resolved. Of note, this limitation does not apply to simulations with neo-Hookean elasticity where, as demonstrated here, oscillations exist with comparable wavelength for different choices of degradation functions $g(\phi)$ and $w(\phi)$ (including those of the model with reduced lattice pinning and those of the KKL model). This somewhat alleviates doubts on the use of a phenomenological description of failure processes inside the dissipation zone, inherent in a phase-field approach, to investigate dynamic fracture instabilities.

Second, the results of the simulations with near tip nonlinear neo-Hookean elasticity support the existence of a supercritical Hopf bifurcation, as evidenced by the fact that the oscillation amplitude $A$ increases rapidly and monotonously as a function of crack velocity $v$ for $v\!>\!v_{\rm c}$. While our simulations lack the resolution to quantitatively demonstrate the scaling $A\!\sim\!\sqrt{v-v_{\rm c}}$ --- theoretically expected for such a bifurcation ---, we do not observe the type of hysteretic behavior that would point to a subcritical bifurcation, at least as far as purely oscillatory behavior is concerned.

Third, the results further support the universal character of the nonlinear oscillatory instability by showing that it is ostensibly independent of $\beta$. An exhaustive series of simulations for different values of $\beta$, varying by an order of magnitude, encompassing regimes where $\Gamma(v)$ is weakly and strongly dependent on $v$, reveal that $\beta$ only has a weak effect on the oscillatory instability wavelength. This finding is consistent with the theoretical expectation that this wavelength is predominantly determined by elastic nonlinearity through the length scale $\ell$ that is independent of the rate of energy dissipation. This result could potentially be tested experimentally in thin brittle materials where $\Gamma$ is velocity independent and $\xi$ is comparable to the sample thickness, so as to suppress 3D micro-branching. Whether such a material can be found is unclear as ideally brittle materials such as glass typically have a very small process zone.

Fourth, the simulation results shed light on the physical origin of the tip-splitting instability by showing that its onset velocity is affected by the wave-speeds inside the dissipation zone, which can be varied phenomenologically by varying the degree of degradation of the kinetic energy inside that zone. In the limit of no degradation, where the wave-speeds drop due to the degradation of the elastic moduli inside that zone, the velocity of straight cracks is limited to about half $c_s$ or less, consistent with previous findings~\citep{Karma2004,Henry.08}. Tip-splitting becomes inevitable as a direct consequence of the limited rate of energy transport and occurs symmetrically, i.e.~with the main crack splitting symmetrically into two branches with equal angles with respect to the parent crack propagation axis. In the opposite limit, where the kinetic energy is fully degraded, this degradation compensates the degradation of the moduli so as to keep the wave-speeds constant inside the entire dissipation zone and straight crack propagation is only limited by the wave-speeds. In this case, tip-splitting still occurs above a critical velocity that is very close to the one corresponding to the onset of oscillations. As a result, for $v$ slightly above $v_{\rm c}$, tip-splitting can occur asymmetrically from an oscillatory crack state, manifested as the emission of a side branch that is somewhat reminiscent of 3D micro-branching.

Extension of the present simulations to 3D are presently underway to investigate the tantalizing possibility that this asymmetric form of tip-splitting is related to micro-branching in 3D. This possibility is suggested by experiments showing that this type of side-branching can be induced to occur for $v\!<\!v_c$ with a finite mode-II perturbation with an amplitude that becomes vanishing small as $v\!\rightarrow\!v_{\rm c}$~\citep{boue2015origin}.

\vspace{0.5cm}

\noindent{\bf Acknowledgements}\\

This research was supported by a grant from the United States-Israel Binational Science Foundation
(BSF, Grant No.~2018603), Jerusalem, Israel, and the United States National Science Foundation (NSF, Grant No.~1827343).
E.B.~also acknowledges support from the Ben May Center for Chemical Theory and Computation and the Harold Perlman Family.

\vspace{0.5cm}

\appendix

\noindent{\bf Appendix}
\setcounter{figure}{0}

\section{Numerical discretization scheme}
\label{sec:AppendixA}

The goal of this appendix is to provide a detailed description of the numerical discretization scheme of the equations of motion, cf.~Eqs.~\eqref{eq:eom-phi}-\eqref{eq:eom-uy}. The latter, using Eq.~\eqref{Eq:Lagrangian}, can be presented as
\begin{align}
\frac{1}{\chi}\frac{\partial \phi}{\partial t} &= \kappa \nabla^2 \phi - g'(\phi) e_{\mbox{\scriptsize{strain}}} - w'(\phi) e_{\rm c} + \frac{1}{2}\rho\,f'(\phi)\bm{v}\cdot \bm{v} \ , \label{Eq:eom-phiApp}\\
\frac{\partial{\bm{u}}}{\partial t} &= \bm{v} \ ,\\
\rho\,f(\phi)\,\frac{\partial{\bm{v}}}{\partial t} &=\nabla {\cdot}{\bm{P}} -\rho\,\frac{\partial f(\phi)}{\partial t}\bm{v} \ .\label{Eq:eom-uApp}
\end{align}
Here ${\bm P}$ is the first Piola-Kirchoff stress tensor, defined as the stress that is thermodynamically conjugate to ${\bm{F}}$~\citep{Holzapfel}, given by
\begin{equation}
\bm{P}=g(\phi)\frac{\partial e_{\text{strain}}\left(\bm{F}\right)}{\partial\bm{F}} \ ,
\label{Eq:P_def_Appendix}
\end{equation}
and the operator $\nabla {\cdot}$ is the divergence operator with respect to the undeformed coordinates, defined as
\begin{equation}
\left(\nabla\cdot\bm{P}\right)_{i}=\partial_{j}P_{ij} \ .
\label{Eq:div_tensor_Appendix}
\end{equation}
The degradation functions $g(\phi)$, $w(\phi)$ and $f(\phi)$ are assumed to be given, as well as the elastic energy density functional $e_{\mbox{\scriptsize{strain}}}$. For the latter, we use in this paper either Eq.~\eqref{eq:neoHookean} or its linear elastic approximation $e^{\mbox{\tiny{le}}}_{\mbox{\scriptsize{strain}}}$ (see main text for exact definition). We aim at numerically calculating the phase field $\phi(x,y,t)$, the displacement vector field ${\bm{u}}(x,y,t)$ and the velocity field $\bm{v}(x,y,t)$. Consequently, Eqs.~\eqref{Eq:eom-phiApp}-\eqref{Eq:eom-uApp} are discretized in both space and time, as detailed next. We will first outline the discretization of these equations in space and then discuss the discretization in time.
\begin{figure}[h]
\includegraphics[height=2.5 in]{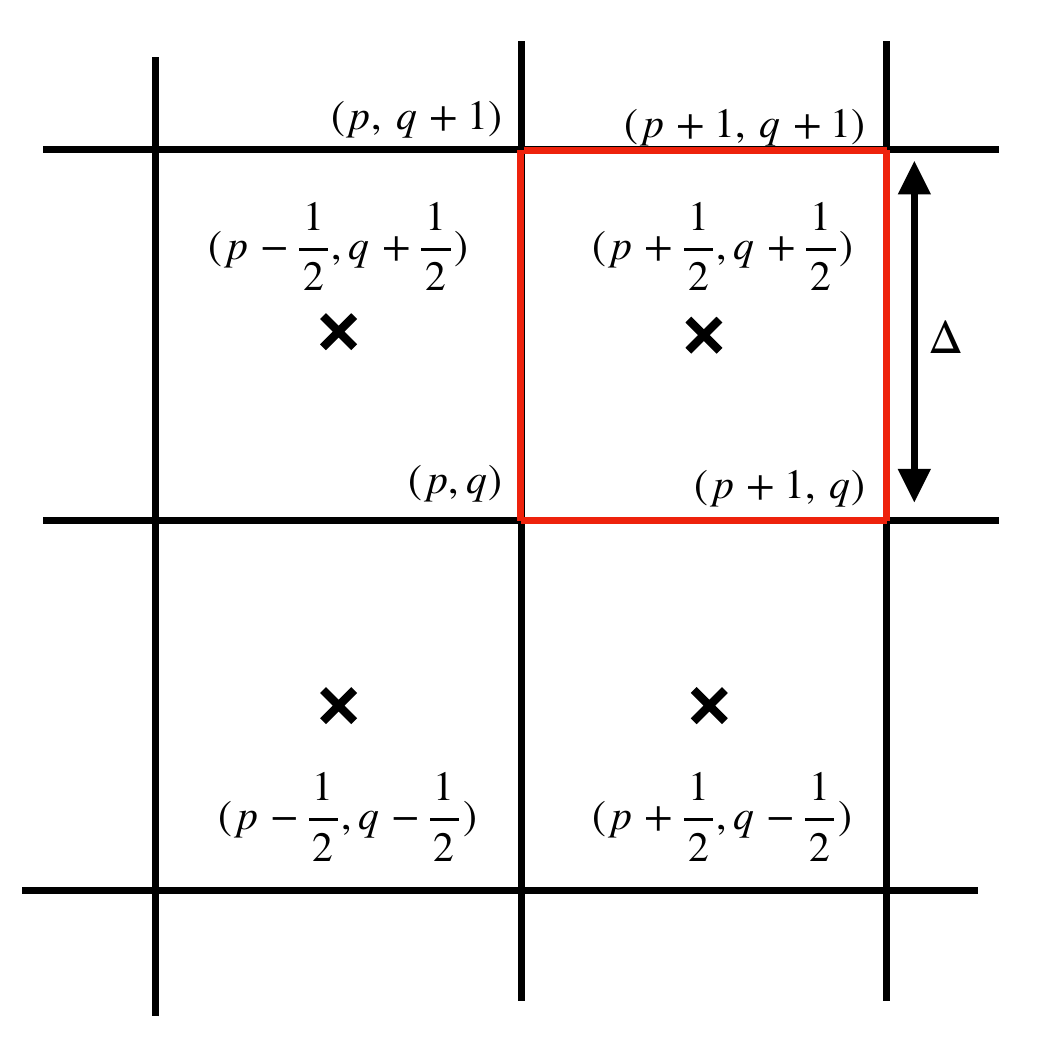}
\centering
\caption{The spatial discretization of the domain into a uniform square grid with spacing $\Delta$. Shown in red is a representative element with coordinates  $(p\Delta,q\Delta), ((p+1)\Delta,q\Delta), ((p+1)\Delta,(q+1)\Delta), (p\Delta,(q+1)\Delta)$.}
\label{Fig:discretization}
\end{figure}

\subsection{Spatial discretization}
\label{subsec:spatial_discretization}

Equations~\eqref{Eq:eom-phiApp}-\eqref{Eq:eom-uApp} are discretized on a uniform square grid of spacing $\Delta\=0.2\xi$, comprising of $n_x$ points in the $x$-direction and $n_y$ points in the $y$-direction, as shown in Fig.~\ref{Fig:discretization}. The fields $\phi(x,y,t)$, $\bm{u}(x,y,t)$ and $\bm{v}(x,y,t)$ are discretized on the vertices of the grid, denoted by indices $(p,q)$, where $p\!\in\!\{0,1,..n_x-1\}$ and $q\!\in\!\{0,1,..n_y-1\}$. An element of the grid with corners $ (p,q), (p,q+1), (p+1,q) $ and $ (p+1, q+1)$ is represented by the indices of its center, i.e.~$(p + \frac{1}{2}, q + \frac{1}{2})$. The components of the deformation gradient tensor, $F_{ij}$ with $i,j\!=\!\{x,y\}$,  are approximated at the center of each element as
\begin{subequations}
\label{Eq:F_derive_Appendix}
\begin{align}
F_{\rm{xx}}^{\left( p + \frac{1}{2}, q + \frac{1}{2}\right)} &= 1 + \frac{\partial u_{\rm{x}}}{\partial x}^{\left( p + \frac{1}{2}, q + \frac{1}{2}\right)} = 1 + \frac{1}{2} \left[ \frac{\partial u_{\rm{x}}}{\partial x}^{\left( p + \frac{1}{2}, q \right)} + \frac{\partial u_{\rm{x}}}{\partial x}^{\left( p + \frac{1}{2}, q + 1 \right)}\right] \\
&= 1 + \frac{1}{2\Delta} \left[ (u_{\rm{x}}^{\left( p + 1, q \right)} - u_{\rm{x}}^{\left( p, q \right)}) + (u_{\rm{x}}^{\left( p + 1, q +1\right)} - u_{\rm{x}}^{\left( p, q +1\right)}) \right] \nonumber \ ,\\
F_{\rm{xy}}^{\left( p + \frac{1}{2}, q + \frac{1}{2}\right)} &= \frac{1}{2\Delta} \left[ (u_{\rm{x}}^{\left( p, q +1\right)} - u_{\rm{x}}^{\left( p, q \right)}) + (u_{\rm{x}}^{\left( p + 1, q +1\right)} - u_{\rm{x}}^{\left( p+1, q \right)}) \right] \ ,\\
F_{\rm{yx}}^{\left( p + \frac{1}{2}, q + \frac{1}{2}\right)} &= \frac{1}{2\Delta} \left[ (u_{\rm{y}}^{\left( p+1, q \right)} - u_{\rm{y}}^{\left( p, q \right)}) + (u_{\rm{y}}^{\left( p + 1, q +1\right)} - u_{\rm{y}}^{\left( p, q +1 \right)}) \right] \ , \\
F_{\rm{yy}}^{\left( p + \frac{1}{2}, q + \frac{1}{2}\right)} &= 1 + \frac{1}{2\Delta} \left[ (u_{\rm{y}}^{\left( p, q +1\right)} - u_{\rm{y}}^{\left( p, q \right)}) + (u_{\rm{y}}^{\left( p + 1, q +1\right)} - u_{\rm{y}}^{\left( p+1, q \right)}) \right] \ .
\end{align}
\end{subequations}
$\bm{P}$ is evaluated at center of the elements, at the points $(p+\frac{1}{2},q+\frac{1}{2})$, as
\begin{equation}
\bm{P}^{\left(p+\frac{1}{2},q+\frac{1}{2}\right)}=g\left(\phi\right)^{\left(p+\frac{1}{2},q+\frac{1}{2}\right)}\frac{\partial e_{\text{strain}}\left(\bm{F}^{\left(p+\frac{1}{2},q+\frac{1}{2}\right)}\right)}{\partial\bm{F}}  \ ,
\label{Eq:P_descrit_Appendix}
\end{equation}
where $g\left(\phi\right)^{\left(p+\frac{1}{2},q+\frac{1}{2}\right)}$ is the approximated as the average of $g(\phi)$ at the neighboring vertices, evaluated as
\begin{equation}
g(\phi)^{(p+\frac{1}{2},q+\frac{1}{2})}=
\frac{1}{4}\left(g(\phi^{(p,q)})+g(\phi^{(p+1,q)})+g(\phi^{(p,q+1)})+g(\phi^{(p+1,q+1)})\right) \ .
\label{Eq:mg_Appendix}
\end{equation}
The strain energy density at the vertex $(p,q)$ is approximated as the average of the strain energy densities evaluated at the centers of the neighboring elements, expressed as
\begin{equation}
e^{(p,q)}_{\text{strain}}=\frac{1}{4}\left(e_{\text{strain}}^{\left(p+\frac{1}{2},q+\frac{1}{2}\right)}+e_{\text{strain}}^{\left(p+\frac{1}{2},q-\frac{1}{2}\right)}
+e_{\text{strain}}^{\left(p-\frac{1}{2},q+\frac{1}{2}\right)}+e_{\text{strain}}^{\left(p-\frac{1}{2},q-\frac{1}{2}\right)}\right) \ .
\label{Eq:e_strain_Appendix}
\end{equation}
The numerical approximation of the Laplacian of the phase-field is given as
\begin{align}
\kappa\nabla^{2}\phi^{\left(p,q\right)} & =\kappa\left(\frac{\partial^{2}\phi^{\left(p,q\right)}}{\partial x^{2}}+\frac{\partial^{2}\phi^{\left(p,q\right)}}{\partial y^{2}}\right)\nonumber \\
 & =\frac{\kappa}{\Delta^{2}}\left(\phi^{\left(p+1,q\right)}+\phi^{\left(p-1,q\right)}+\phi^{\left(p,q+1\right)}+\phi^{\left(p,q-1\right)}-4\phi^{\left(p,q\right)}\right) \ .
\label{Eq:lap_phi_Appendix}
\end{align}
Finally, $(\nabla\cdot\bm{P})$ is evaluated at the points $\left(p,q\right)$, similarly to Eqs.~\eqref{Eq:F_derive_Appendix}, as
\begin{subequations}
\begin{align}
\label{Eq:div_P_Appendix}
\left(\nabla\cdot\bm{P}\right)_{\text{x}}^{\left(p,q\right)} & =\frac{1}{2\Delta}\left(P_{\text{xx}}^{\left(p+\frac{1}{2},q+\frac{1}{2}\right)}-P_{\text{xx}}^{\left(p-\frac{1}{2},q+\frac{1}{2}\right)}+P_{\text{xx}}^{\left(p+\frac{1}{2},q-\frac{1}{2}\right)}-P_{\text{xx}}^{\left(p-\frac{1}{2},q-\frac{1}{2}\right)}\right)\\
 & +\frac{1}{2\Delta}\left(P_{\text{xy}}^{\left(p+\frac{1}{2},q+\frac{1}{2}\right)}-P_{\text{xy}}^{\left(p+\frac{1}{2},q-\frac{1}{2}\right)}+P_{\text{xy}}^{\left(p-\frac{1}{2},q+\frac{1}{2}\right)}-P_{\text{xy}}^{\left(p-\frac{1}{2},q-\frac{1}{2}\right)}\right)\nonumber \ ,\\
\left(\nabla\cdot\bm{P}\right)_{\text{y}}^{\left(p,q\right)} & =\frac{1}{2\Delta}\left(P_{\text{yx}}^{\left(p+\frac{1}{2},q+\frac{1}{2}\right)}-P_{\text{yx}}^{\left(p-\frac{1}{2},q+\frac{1}{2}\right)}+P_{\text{yx}}^{\left(p+\frac{1}{2},q-\frac{1}{2}\right)}-P_{\text{yx}}^{\left(p-\frac{1}{2},q-\frac{1}{2}\right)}\right)\\
 & +\frac{1}{2\Delta}\left(P_{\text{yy}}^{\left(p+\frac{1}{2},q+\frac{1}{2}\right)}-P_{\text{yy}}^{\left(p+\frac{1}{2},q-\frac{1}{2}\right)}+P_{\text{yy}}^{\left(p-\frac{1}{2},q+\frac{1}{2}\right)}-P_{\text{yy}}^{\left(p-\frac{1}{2},q-\frac{1}{2}\right)}\right)\nonumber\ .
\end{align}
\end{subequations}
Using the spatial discretization described above, the equations of motion can be rewritten as
\begin{align}
\frac{1}{\chi}\frac{\partial \phi ^{\left( p,q\right)} (t)}{\partial t} &= \mathcal{G}^{\left( p,q\right)}\left(\bm{u}(t),\bm{v}^{\left( p,q\right)}(t),\phi^{\left( p,q\right)}(t)\right) \label{Eq:phi-t}\ ,\\
\frac{\partial\bm{u}^{\left( p,q\right)}}{\partial t}(t)=\bm{v}^{\left( p,q\right)}(t)\ ,\\
\rho\,f(\phi^{(p,q)})\,\frac{\partial{\bm{v}}^{\left( p,q\right)}(t)}{\partial t} &=\mathcal{H}^{\left( p,q\right)}\left(\bm{u}(t),\bm{v}^{\left( p,q\right)}(t),\phi^{\left( p,q\right)}(t)\right)\label{Eq:u-t}\ ,
\end{align}
where
\begin{align}
\mathcal{G} ^{\left( p,q\right)} = \kappa\nabla^{2}\phi^{\left(p,q\right)} - g'(\phi^{(p,q)}) e_{\mbox{\scriptsize{strain}}}^{\left( p,q\right)} - w'(\phi^{(p,q)}) e_{\rm c} +
& \frac{1}{2}\,\rho\,f'(\phi^{(p,q)})\,\bm{v}^{(p,q)} \cdot \bm{v}^{(p,q)} \nonumber\ ,
\end{align}
and
\begin{align}
\mathcal{H} ^{\left( p,q\right)} = \left(\nabla \cdot{\bm{P}}\right)^{(p,q)} -\rho\,\frac{\partial f^{(p,q)}}{\partial t} \bm{v}^{(p,q)} \label{Eq:H} \nonumber \ .
\end{align}
Note that ${\bm u}$ in the argument of $\mathcal{G}^{\left( p,q\right)}$ in Eq.~\eqref{Eq:phi-t} and of $\mathcal{H}^{\left( p,q\right)}$ in Eq.~\eqref{Eq:u-t} does not carry a discretization index as the dependence on ${\bm u}$ is {\em nonlocal}.

\subsection{Temporal discretization}
\label{subsec:temporal_discretization}

The phase-field equation, i.e.~Eq.~\eqref{Eq:phi-t}, is discretized using a simple forward Euler scheme
\begin{equation}
\phi^{(p,q)}_{n+1} = \phi^{(p,q)}_{n} + \mathcal{G}^{\left(p,q\right)}\left(\bm{u}_{n},\bm{v}^{(p,q)}_{n},\phi^{(p,q)}_{n}\right) \rm{\Delta t} \ ,
\end{equation}
where the subscript $n$ refers to the current time step, $ t_n = n \Delta t$, where $ \rm{\Delta t}$ is the time increment. The time evolution of the displacement is computed using a modified Beeman's algorithm~\citep{schofield1973computer,beeman1976some,LEVITT1983617}, according to
\begin{align}
\bm{a}_{n}^{\left(p,q\right)} & =\frac{1}{\rho\,f\left(\phi_{n+1}^{\left(p,q\right)}\right)}\left[\left(\nabla\cdot\bm{P}\left(\bm{u}_{n},\phi_{n+1}\right)\right)^{\left(p,q\right)}-\rho\left(\frac{f\left(\phi_{n+1}\right)-f\left(\phi_{n}\right)}{\Delta t}\right)\bm{v}_{n}^{\left(p,q\right)}\right]\ ,\\
\bm{u}_{n+1}^{\left(p,q\right)} & =\bm{u}_{n}^{\left(p,q\right)}+\bm{v}_{n}^{\left(p,q\right)}\Delta t+\frac{1}{6}\left(4\bm{a}_{n}^{\left(p,q\right)}-\bm{a}_{n-1}^{\left(p,q\right)}\right)\Delta t^{2}\ ,\\
\tilde{\bm{a}}_{n+1}^{\left(p,q\right)} & =\frac{1}{\rho\,f\left(\phi_{n+1}^{\left(p,q\right)}\right)}\left[\left(\nabla\cdot\bm{P}\left(\bm{u}_{n+1},\phi_{n+1}\right)\right)^{\left(p,q\right)}-\rho\left(\frac{f\left(\phi_{n+1}\right)-f\left(\phi_{n}\right)}{\Delta t}\right)\bm{v}_{n}^{\left(p,q\right)}\right]\ ,\\
\bm{v}_{n+1}^{\left(p,q\right)} & =\bm{v}_{n}^{\left(p,q\right)}+\frac{1}{12}\left(5\tilde{\bm{a}}_{n+1}^{\left(p,q\right)}+8\bm{a}_{n}^{\left(p,q\right)}-\bm{a}_{n-1}^{\left(p,q\right)}\right)\Delta t \ .
\end{align}

\subsection{Numerical regularization of the strain energy density}
\label{subsec:e_regularization}

Under mode-I loading conditions, the material is under tension and hence the out-of-plane stretch ratio $[\det({\bm {F}})]^{-1}$ appearing in the strain energy density  in Eq.~\eqref{eq:neoHookean} is smaller than unity everywhere in space. However, in dynamic situations such as those encountered in fracture, transient nonlinear elastic waves can cause compression behind the crack tip, leading to very large $[\det({\bm {F}})]^{-1}$. To avoid numerical issues associated with such large values, which are not expected to affect the properties of our solutions, we modify the strain energy density in Eq.~\eqref{eq:neoHookean} to read~\citep{Chen2017}
\begin{eqnarray}
e_{\mbox{\scriptsize{strain}}}=\frac{\mu}{2}\,\left(F_{ij}F_{ij}+\frac{1}{J^2}-3\right) \ , \label{eq:e_strainReg}
\end{eqnarray}
where
\begin{equation}
\frac{1}{J^2}\equiv
\begin{dcases}
1/[\det({\bm {F}})]^{2},& \text{if } [\det({\bm {F}})]^{-1}< J_{\rm{min}}^{-1}\\
8/[\det({\bm {F}}) + J_{\rm{min}}]^2 - 1/J_{\rm{min}}^2, & \text{otherwise} \ ,
\end{dcases}
\label{Eq:Jmin}
\end{equation}
with $J_{\rm{min}}$ being a numerical cutoff parameter. Equation~\eqref{Eq:Jmin} regularizes the strain energy density function $e_{\rm{strain}}$ in the $\det({\bm {F}})\!\to\!0$ limit and ensures the continuity of the first derivative of $e_{\rm{strain}}$. In our simulations, we chose $J_{\rm{min}}\!=\!0.2$ and found that this choice has a negligible influence on the crack dynamics; the same applies to other choices of $J_{\rm{min}}$, as long as it is chosen to be much smaller than unity.

\subsection{Simulation setup for the oscillatory and tip-splitting instabilities}
\label{subsec:treadmill}

\begin{figure}[ht!]
\includegraphics[ width=0.9\textwidth]{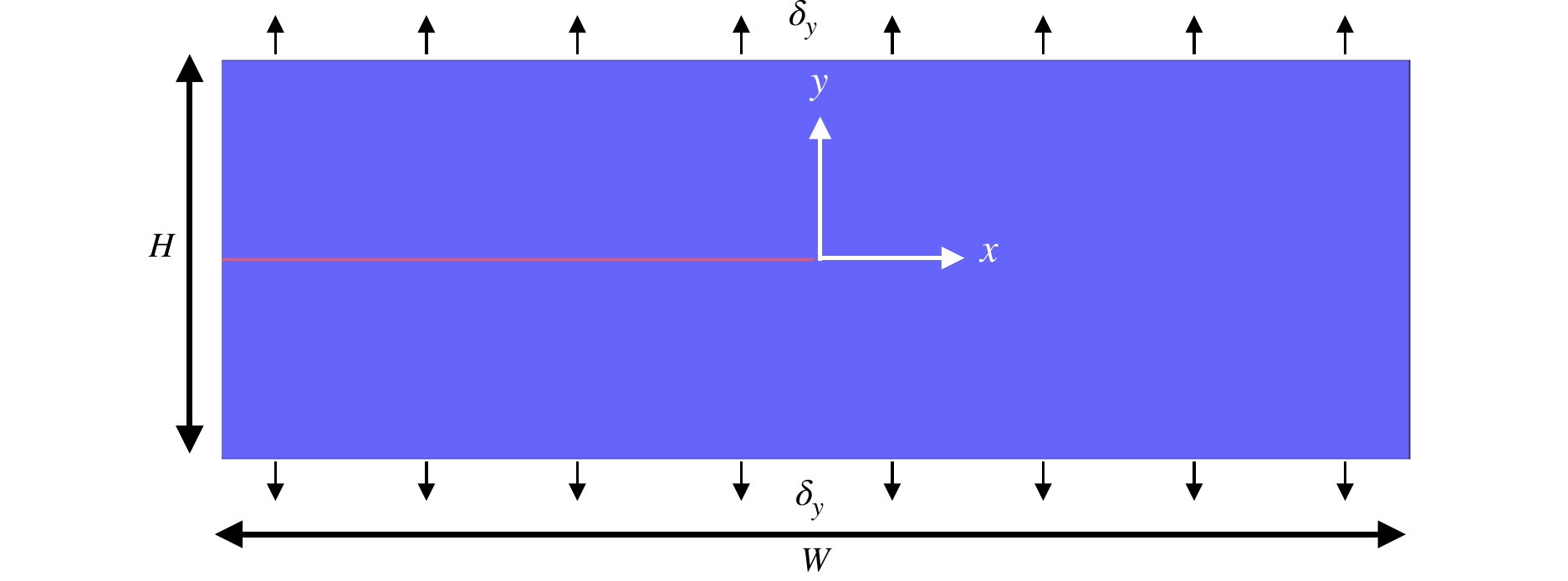}
\centering
\caption{The simulation set up for mode-I dynamic fracture. All symbols are defined in the text.}
\label{Fig:Simulation-setup}
\end{figure}

In Fig.~\ref{Fig:Simulation-setup} we show a typical initial configuration that is used in this work to investigate 2D high-velocity fracture instabilities. A rectangular strip of dimension $H$ (in the $y$-direction) and $W$ (in the $x$-direction) contains an edge crack along the symmetry line in the $x$-direction, which extends up to the center of the strip.

The strip is loaded in pure mode-I by fixing the vertical and horizontal displacement $u_y(y\!=\!\pm H/2)\!=\!\pm\delta_y$ and $u_x(y\!=\!\pm H/2)\=0$ on the top and bottom edges of the strip. Prior to the initiation of the simulation, the displacement field $\bm{u}$ is relaxed to equilibrium, $\nabla\!\cdot\!\boldsymbol{P}\!=\!0$, while keeping $\phi$ fixed. During the relaxation procedure, the boundary conditions far behind the tip, $x\=-W/2$ (left edge of the strip), and far ahead of it, $x\=W/2$ (right edge of the strip), are set to $\partial_{x}u_{x}(x\!=\!-W/2)\!=\!\partial_{x}u_{y}(x\!=\!-W/2)\!=\!\partial_{x}u_{x}(x=W/2)\!=\!\partial_{x}u_{y}(x\!=\!W/2)\!=\!0$.
During crack dynamics, the boundary conditions are set to $\partial_{x}u_{x}(x\!=\!-W/2)\!=\!\partial_{x}u_{y}(x\!=\!-W/2)\!=\!\partial_{x}v_{x}(x\!=\!-W/2)\!=\!\partial_{x}v_{y}(x\!=\!-W/2)\!=\!\partial_{x}\phi(x\!=\!-W/2)\!=\!0$ behind the crack, and to $v_{x}(x\!=\!W/2)\!=\!v_{y}(x\!=\!W/2)\!=\!\partial_{t}\phi(x\!=\!W)/2\!=\!0$ ahead of the crack.

A treadmill procedure is used to simulate a strip of effectively infinite length, where a strained layer is added on the right vertical boundary while another layer is removed from the opposite left boundary, such that the crack tip always remains at the center of the strip. This allows to propagate the crack for very large distances, with negligible boundary effects. A small amount of Kelvin's dissipation is added behind the crack, or close to the system boundaries, to damp the effect of elastic waves generated in dynamic simulations. In a typical simulation, we have $H\=300\xi-800\xi$ and $W/H\=1\!-\!3$, with a spacing of $\Delta\=0.2\xi$, resulting in a system with $10^6\!-\!10^7$ degrees of freedom. A time step of $\Delta t\=8\!\times\!10^{-4}$ is used and the simulation codes are parallelized on NVIDIA GPU's using the CUDA platform.

\section{Numerical evaluation of the J-integral}
\label{sec:AppendixB}
\setcounter{figure}{0}

\begin{figure}[ht!]
\includegraphics[width=0.9\textwidth]{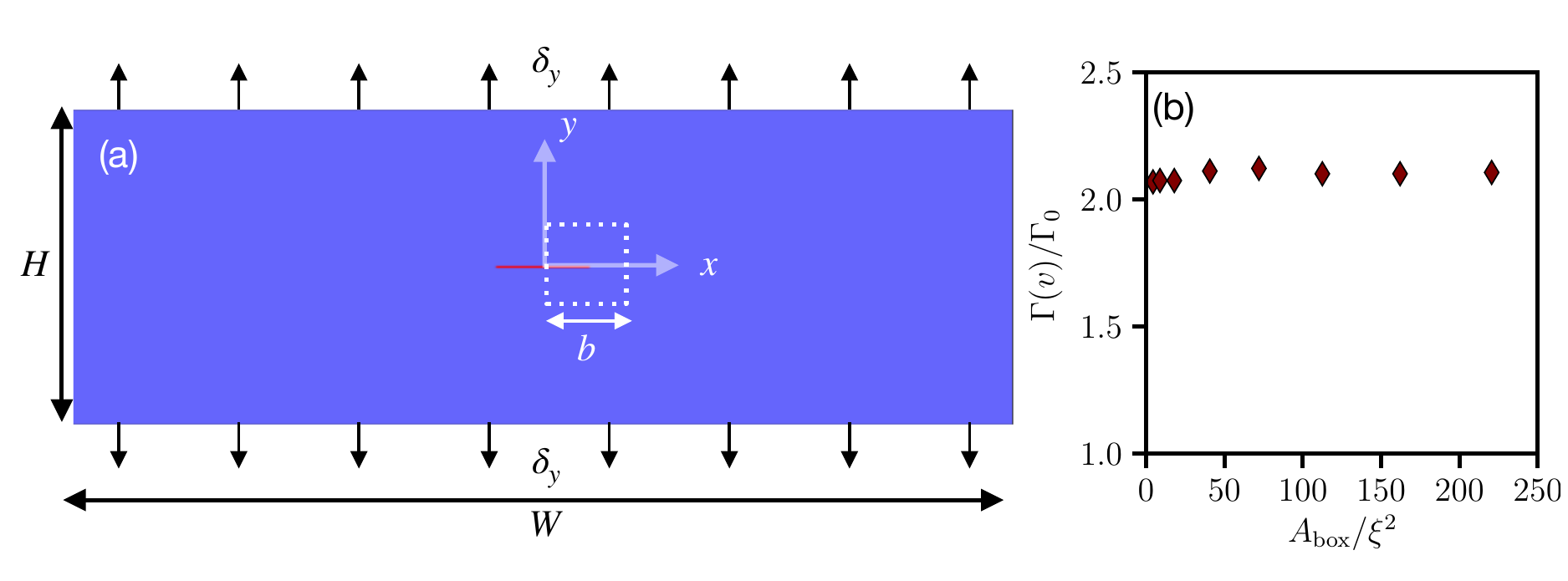}
\centering
\caption{(a) The simulation set up used for the calculation of the J-integral, see text for details. (b) The normalized fracture energy $\Gamma(v)/\Gamma_0$, measured using the J-integral for a crack propagating at an instantaneous velocity $v/c_s\!=\!0.79$, as a function the area enclosed in the contour box, $A_{\rm{box}}\!=\!b^2$, where $b$ is the linear size of the contour box, as shown in panel (a). Here we used the linear elastic strain energy density $e^{\mbox{\tiny{le}}}_{\mbox{\scriptsize{strain}}}$, together with $H\!=\!300\xi$, $W\!=\!900\xi$, $\Delta\!=\!0.2\xi$, $\beta\!=\!2.8$, ${\cal W}/\Gamma_0\!=\!3.0$ and $e_{\rm c}/\mu\!=\!0.5$.}
\label{Fig:Simulation-setup-J-Integral}
\end{figure}

In order to numerically evaluate the J-integral, the simulation setup is modified to have a seed crack located at the center along the symmetry line in the $x$-direction. The crack then propagates outwards in both the positive and negative $x$-directions. Vertical displacements $u_y(\pm H/2)\=\pm\delta_y $ are applied to the top and bottom edges of the strip and no-flux boundary conditions are applied to the right and left vertical boundaries,  i.e.~$\partial_x u_x, \partial_x u_y, \partial_x \phi$ are all set to zero. To evaluate the J-integral, see explicit expression in the text, the integral is evaluated on a contour surrounding the crack tip that is chosen as a square box of linear size $b$, cf.~Fig.~\ref{Fig:Simulation-setup-J-Integral}a. The J-integral is computed for different box sizes $b$ and the result is shown to be independent of $b$, cf.~Fig.~\ref{Fig:Simulation-setup-J-Integral}b.

\numberwithin{equation}{section}

\newpage

\section{The near tip fields of straight cracks propagating at high velocities}
\label{subsec:tip_fields}\label{sec:AppendixC}
\setcounter{figure}{0}

The fully dynamic phase-field approach allows to quantitatively address various basic aspects of fast crack propagation. For example, in~\citet{Bouchbinder.14} (cf.~Eq.~(49) therein) it has been shown that the singular $\sim\!1/\sqrt{r}$ mode-I LEFM contribution to the extensional strain $\epsilon_{yy}\=\partial_{y}u_{y}$ ahead of a propagating crack can become \emph{negative}; this happens if the crack velocity satisfies $v/c_s\!>\!\tfrac{1}{2}\Big(\!\sqrt{(c_d/c_s)^2+8}-c_d/c_s\!\Big)$, with $c_d$ being the dilatational wave-speed. As the singular contribution is expected to dominate $\epsilon_{yy}$ over some spatial range, we expect that for $v/c_s\!>\!\tfrac{1}{2}\Big(\!\sqrt{(c_d/c_s)^2+8}-c_d/c_s\!\Big)$ one observes $\epsilon_{yy}(x,y\=0)\!<\!0$ at some intermediate range of $x$'s  ahead of the propagating tip. It is clear that mode-I fracture is driven by extensional strains, so we also have $\epsilon_{yy}\!>\!0$ far enough ahead of the tip, i.e.~for sufficiently large $x$. Recall that $x\=0$ is the crack tip location, cf.~Fig.~\ref{Fig:Simulation-setup}.

While $\epsilon_{yy}(x,y\=0)\!<\!0$ might appear physically inconsistent, as mode-I tensile fracture is ultimately related to extensional (opening) strains, it contradicts nothing. The existence of $\epsilon_{yy}(x,y\=0)\!<\!0$ over some range of $x$'s ahead of a propagating crack tip simply implies that at yet smaller $x$'s, $\epsilon_{yy}(x,y\=0)$ should change sign again and become positive where material failure is actually taking place. In the absence of near tip elastic nonlinearity, i.e.~for $\ell\=0$, the intervention of the dissipation length is expected to be responsible for $\epsilon_{yy}(x,y\=0)$ becoming positive again. Consequently, by continuity, this implies the existence of a region larger than the dissipation zone where significant deviations from the LEFM singular fields are expected under strongly dynamic conditions. This deviation sets a dynamic length scale that is associated with the presence of a finite dissipation zone and of tip blunting (cf.~Fig.~\ref{Fig:f-phi}).

\begin{figure}[ht!]
\includegraphics[width=0.6\textwidth]{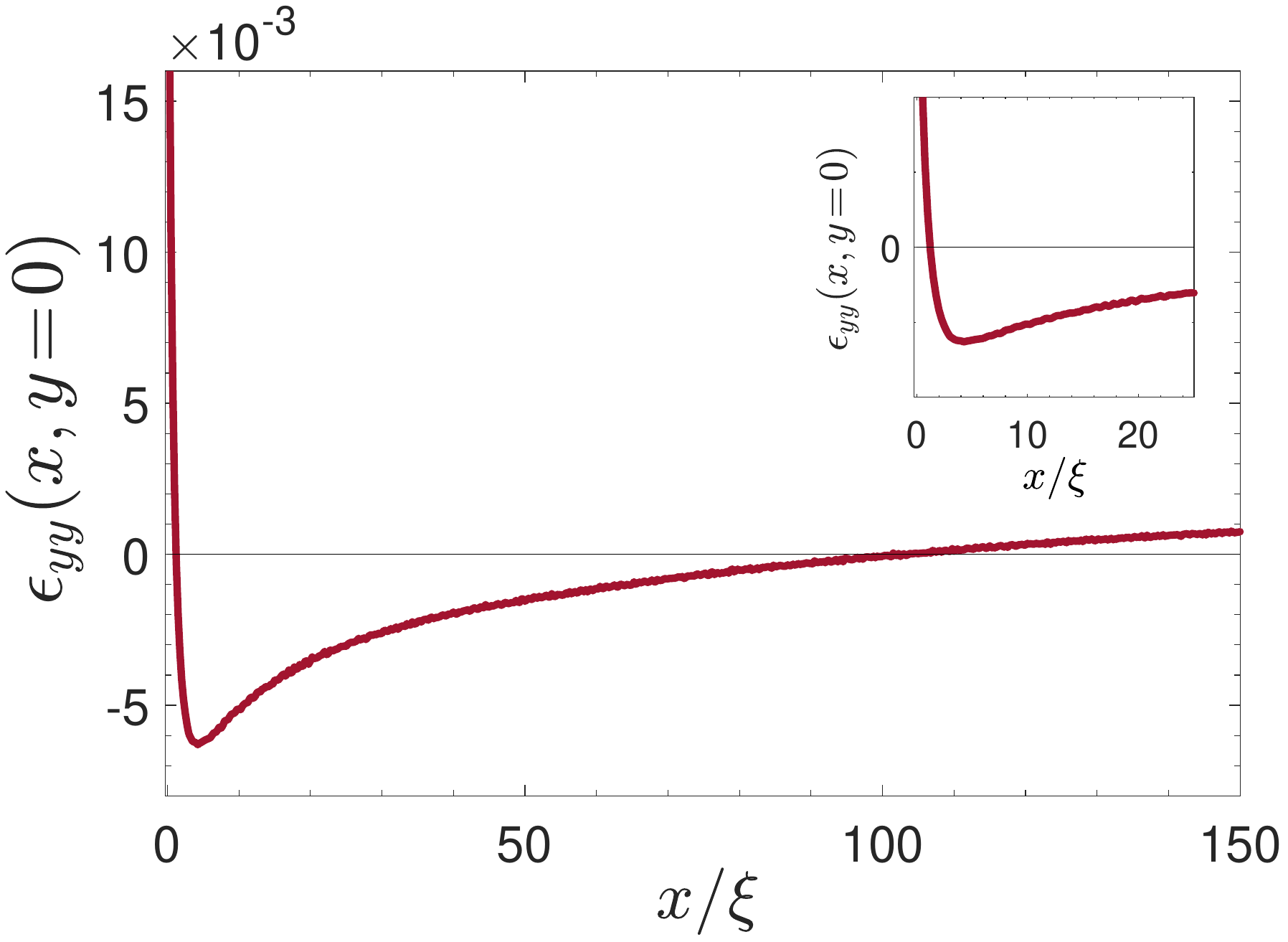}
\centering
\caption{The extensional strain $\epsilon_{yy}(x,y\!=\!0)\!=\!\partial_{y}u_{y}(x,y\!=\!0)$ as a function of $x/\xi$ for a mode-I crack propagating at $v\!=\!0.87c_s$. The system size used is $H\!=\!W\!=\!2000\xi$ (cf.~Fig.~\ref{Fig:Simulation-setup}), and we set $\ell\!=\!0$ (i.e.~no near tip elastic nonlinearity) and $c_d\!=\!2c_s$. Under these conditions (see text for discussion), it is theoretically predicted that $\epsilon_{yy}(x,y\!=\!0)$ is nonmonotonic and negative at intermediate $x$'s, exactly as observed. $\epsilon_{yy}(x,y\!=\!0)$ attains a minimum at $x\!\simeq\!4.5\xi$, indicating that the magnitude of the zone in which LEFM breaks down at high propagation velocities is~of ${\cal O}(10\xi)$, due to dynamic renormalization effects. (inset) Zooming in on the minimum region.}
\label{Fig:tip_fields}
\end{figure}

In Fig.~\ref{Fig:tip_fields} we present $\epsilon_{yy}(x,y\=0)$ for a mode-I crack propagating at $v\=0.87c_s$ with $\ell\=0$ (i.e.~no near tip elastic nonlinearity exists in this case) and $c_d\=2c_s$. Since for the latter we have $\tfrac{1}{2}\Big(\!\sqrt{(c_d/c_s)^2+8}-c_d/c_s\!\Big)\=0.73\!<\!0.87$, we expect $\epsilon_{yy}(x,y\=0)$ to follow the predictions just discussed. Indeed, these predictions are fully verified in Fig.~\ref{Fig:tip_fields}, where $\epsilon_{yy}(x,y\=0)$ is observed to change sign from positive to negative and then to positive again with decreasing $x$. The minimum of $\epsilon_{yy}(x,y\=0)$ provides a lower bound on the magnitude of the zone where the singular $\sim\!1/\sqrt{r}$ fields are not dominant anymore (since deviations from the singular $\sim\!1/\sqrt{r}$ fields must occur even before the minimum is reached).

The minimum of $\epsilon_{yy}(x,y\=0)$ in Fig.~\ref{Fig:tip_fields} is attained at $x\!\simeq\!4.5\xi$, which suggests that the magnitude of the zone where the singular $\sim\!1/\sqrt{r}$ fields are not dominant anymore, at this high propagation velocity, is of ${\cal O}(10\xi)$. As the onset of the oscillatory instability takes place at a slightly larger propagation velocity (around $0.9c_s$, cf.~Fig.~\ref{fig:LEPF-panel}c), the suggestion that the magnitude of the zone in which LEFM breaks down at high propagation velocities (due to dynamic renormalization effects) can be quite significantly larger than $\xi$ --- i.e.~of ${\cal O}(10\xi)$ --- appears to be consistent with the observation of~Fig.~\ref{fig:LEPF-panel}b in which the oscillations wavelength for $\ell\=0$ is $\lambda\!\simeq\!13\xi$.

\numberwithin{equation}{section}


\end{document}